\documentclass[aps,physrev,reprint,groupedaddress,showkeys]{revtex4-2}
\usepackage{xcolor}
\usepackage{tikz}
\usepackage{graphicx}
\usepackage{subcaption}
\usepackage[skip=5pt, font=footnotesize]{caption}
\usepackage{comment}
\usepackage{amsmath}
\usepackage{natbib}
\usepackage{dcolumn}
\usepackage[normalem]{ulem}

\begin{document}

\title{Multiscale complexity of two-dimensional Ising systems with short-range, ferromagnetic interactions}

\author{Ibrahim Al-Azki}
\email[Contact author: ]{ibrahim.al.azki@rmit.edu.au}
\affiliation{Department of Physics, School of Science, RMIT University, Melbourne, Victoria 3000, Australia}

\author{Valentina Baccetti}
\email[Contact author: ]{valentina.baccetti@rmit.edu.au}
\affiliation{Department of Physics, School of Science, RMIT University, Melbourne, Victoria 3000, Australia}

\date{\today}

\begin{abstract}
Complex systems exhibit macroscopic behaviors that emerge from the coordinated interactions of their individual components. Understanding the microscopic origins of these emergent properties remains a significant challenge, especially in less-understood systems, due to the absence of a generalized framework for identifying the governing degrees of freedom. The multiscale complexity formalism, developed to address this challenge, consists of a set of information-theoretic indices designed to identify the scales at which collective behaviors emerge. In this article, we evaluate one such index, the complexity profile, by applying it to the two-dimensional Ising model with finite-range interactions. In particular, we show that the complexity profile captures the transition between the disordered and ordered phases by detecting the emergence of multiscale structure exclusively in the critical region, and therefore offering insights into the formation of magnetic domains from an information-theoretic perspective. Additionally, we show that the pairwise complexity exhibits a maximum in the disordered phase that remains bounded in the thermodynamic limit. These results highlight the potential of the multiscale complexity formalism to probe emergent behaviors and detect hidden features of critical phenomena in interacting systems beyond the classical characterization of correlations
\end{abstract}


\maketitle

\section{Introduction}
Collective behaviors such as brain function and magnetic frustration are strong emergent properties of complex systems that cannot be observed from the properties of their individual components~\cite{bar-yam_mathematical_2004}. Driven by the non-linear interactions between the components, these behaviors arise from the tendency of a system to spontaneously organize into states with macroscopic patterns or order. This process, known as self-organization~\cite{Haken2006}, is typically concomitant with
a continuous phase transition as a system passes a critical point, where long-range
correlations span its entirety. 

Many natural and physical systems exhibit such a process~\cite{RevModPhysMuñoz} and, although intrinsically different, share similar emergent properties~\cite{yaneer2016data}. This observation has driven the development of mathematical models that describe systems with uncertain internal mechanisms by analogy to more familiar ones. While this \textit{ad hoc} modeling approach can be effective, similarity in emergent behaviors does not necessarily imply the same underlying microscopic mechanisms. Bridging the gap between these mechanisms and macroscopic behavior requires deriving the governing mathematical relationships \textit{a priori}, while making minimal assumptions of this kind.

However, uncovering the microscopic mechanisms that govern collective behaviors, along with the effective degrees of freedom (DoF), remains a significant challenge in most complex systems. This challenge arises not only from the absence of a general, system-agnostic mathematical framework but also from the multi-component nature of these systems and the higher-order interactions that often cannot be decomposed into simple combinations of pairwise ones~\cite{Battiston2021}. 

To address this issue, the multiscale complexity formalism~\cite{allen_multiscale_2017} has been proposed as a generalized, model-independent framework for identifying the relevant DoF of a complex system at different \textit{scales}, using two information-theoretic indices of \textit{structure}: the complexity profile (CP) and the marginal utility of information (MUI). Each of these indices captures a different aspect of how the sharing of information among components influences the multiscale organization of a system. 

In this paper, we focus on the CP index, with the aim of investigating whether it can identify such DoF in a self-organizing system. Specifically, we assess its ability to capture the multiscale behavior of the two-dimensional Ising model with nearest-neighbor ferromagnetic interactions. This model is characterized by a well-known transition between disordered (paramagnetic) and ordered (ferromagnetic) phases, as well as long-range spatial correlations in the critical region. 

The CP has previously been shown to effectively capture the transitions of the infinite-range~\cite{gheorghiu-svirschevski_multiscale_2004} and Gaussian~\cite{metzler_multiscale_2005} Ising models, both of which are simplified variants that facilitate analytical treatment of collective behavior in interacting spin systems. In the infinite-range model, each spin interacts equally with all other spins in the system, leading to a mean-field description that exhibits a transition between disordered and ordered phases. In contrast, the Gaussian Ising model does not have a well-defined ordered phase and instead undergoes a stability transition characterized by diverging spin amplitudes when interactions between spins dominate self-interactions. However, both models neglect explicit higher-order spin correlations that, in the short-range model, arise non-trivially from local interactions. This may limit the capacity of the CP to probe the full range of correlations responsible for the emergent behaviors it is intended to capture. In this study, by considering a finite-range model, we investigate the role of these correlations on the behavior of the system near the phase transition.

This notion of multiscale analysis has a long history in the statistical physics of phase transitions. The most prominent example is the renormalization group (RG)~\cite{wilson_renormalization_1983}, which is a conceptual framework developed to identify the DoF that govern the observable properties of a physical system at different length scales. The iterative coarse-graining of local variables reveals the scale-dependent properties of the system and relates the emergent macroscopic behavior to the microscopic DoF. However, the practical execution of an RG transformation is often infeasible for complex physical systems for which identifying a mathematical model or representation is difficult. 

The CP may therefore serve as a generalized tool for characterizing order and symmetries in a complex system due to its formulation in terms of an axiomatic definition of information~\cite{allen_multiscale_2017,bar-yam_multiscale_2004}. While statistical mechanical measures are effective at predicting phase transitions, information-theoretic measures offer a complementary perspective that does not rely on the specific details of a system, or on identifying an order parameter \textit{a priori}. For instance, the mutual information has been widely used to study the properties of classical and quantum spin systems~\cite{stephan_renyi_2010,stephan_shannon_2014,wilms_finite-temperature_2012}. In particular, the Rényi mutual information has been shown to detect a topological phase transition in a two-dimensional (2D) continuous-spin model~\cite{iaconis_detecting_2013}, which is traditionally detected using specialized thermodynamic estimators~\cite{J_M_Kosterlitz_1973}. This highlights the potential of the CP to detect and characterize complex behaviors at different scales, providing a broader perspective than conventional physical observables and mutual information.

This paper is organized as follows: Sec.~\ref{Section: MultiscaleInformationTheory} introduces the core concepts of structure and scale, and outlines the theoretical framework for computing the CP. Sec.~\ref{Section: IsingModel} summarizes key aspects of the 2D Ising model relevant for defining subsystem probability distributions. In Sec.~\ref{Section: ComplexityProfileAnalysis}, we present the results of the CPs for three lattice geometries and relate them to the physical behavior of the model. Sec.~\ref{Section: Discussion} discusses the implications of these results, and we conclude with final remarks in Sec.~\ref{Section: Conclusion}.

\section{Multiscale information theory}\label{Section: MultiscaleInformationTheory}
In this section, we present the theoretical background necessary to compute the CP of the 2D Ising model. We begin by introducing the concepts of dependence, structure and scale, which are fundamental to the multiscale complexity formalism. Then, in Sec.~\ref{Subsection: ComplexityProfile}, we focus on the CP and its definition in terms of Shannon information measures. The subsystem probability distributions required to apply these measures to the Ising model are derived in Sec.~\ref{Section: IsingModel}.

\subsection{Multiscale complexity formalism}\label{Subsection: MultiscaleComplexityFormalism}
The multiscale complexity formalism employs the CP and MUI indices to quantitatively capture the structure of a complex system~\cite{allen_multiscale_2017}. Here, structure is defined as the complete set of dependencies among the components of a system. Each dependency is quantified by the amount of shared information that enables the inference of the state of one component from the state of the others. Since a dependency describes a relationship among components, this amount of information highlights the relevance of its scale, that is, the number of components involved, in characterizing emergent collective behaviors. 

Constraints imposed on the individual behaviors of components by their mutual dependencies give rise to informational redundancies within a system, wherein information describing one subsystem may partially or fully overlap with that of others. These redundancies reflect the extent to which components are jointly constrained and, hence, are closely linked to emergent properties. For example, in systems with long-range correlations, such as spatial Ising systems near the critical region, the state of each component often provides information about the states of others. This leads to coordinated behaviors at multiple spatial or temporal levels of organization, such as synchronization~\cite{Zhang2023}, phase transitions~\cite{PhysRevA.EmergentPhase}, and emergent patterns~\cite{RevModPhys.Photons,munn_multiscale_2024}. Therefore, it is essential to consider the scale at which these redundancies occur to understand how they contribute to emergent behaviors.

In the CP framework, scale refers to the number of components acting in coordination, rather than spatial length, and its precise definition depends on the system under consideration~\cite{bar-yam_mathematical_2004}. In the context of the Ising model, it represents the number of correlated spins that are involved in some collective behavior. This determines the level at which information is shared across different spin subsystems, indicating how detailed or coarse grained an observation is with respect to the system as a whole, rather than localized spatial detail. This differs from the notion of scale in the RG framework, where it refers to spatial resolution and the evolution of physical parameters under successive coarse-graining transformations.

\subsection{Complexity profile: An index of structure}\label{Subsection: ComplexityProfile}
As an index of multiscale structure, the CP quantifies the amount of information shared among \textit{at least} $k$ components, which defines scale, to mathematically capture the emergence of collective behaviors.

To formally define the CP, let $\boldsymbol{\sigma}=\{\sigma_1, \ldots, \sigma_N\}$ represent a system consisting of $N$ components that have the same intrinsic scale, that is, they have identical internal structure. The CP, denoted by $C(k)$, of such systems is then defined as~\cite{bar-yam_multiscale_2004}
\begin{equation}\label{Equation: ComplexityProfile}
    C(k) = \sum_{j=0}^{k-1} (-1)^{k-j-1} \binom{N-j-1}{k-j-1} Q\left(N-j\right),
\end{equation}
where $Q(N-j)$ is the total information describing all subsystems with $N-j$ components. Let $\boldsymbol{\sigma}^{(u)} = \{\sigma_1^{(u)}, \ldots ,\sigma_{N-j}^{(u)}\}$ denote the $u$th such subsystem. For all distinct subsystems $\boldsymbol{\sigma}^{(u)}\subseteq \boldsymbol{\sigma}$ of size $N-j$, the total subsystem information $Q(N-j)$ is computed by summing the contribution of each subsystem as
\begin{equation}\label{Equation: SubsystemInformation}
    Q(N-j) = \sum_{\boldsymbol{\sigma}^{(u)}} H(\sigma_1^{(u)}, \ldots ,\sigma_{N-j}^{(u)}),
\end{equation}
where $H(\sigma_1^{(u)}, \ldots ,\sigma_{N-j}^{(u)})$ is a general information function that quantifies the amount of information needed to describe subsystem $\boldsymbol{\sigma}^{(u)}$. The coefficient of $Q(N-j)$ in Eq.~\eqref{Equation: ComplexityProfile} originates from the inclusion-exclusion principle, which ensures that each dependency measure appears in $C(k)$ with a multiplicity of one. 

While the appropriate information function is chosen depending on some generic properties of a system, it must satisfy the fundamental axioms of monotonicity and subadditivity. The former requires that a smaller subsystem contains no more information than a larger one, while the latter states that the whole cannot contain more information than the sum of the information of its parts (we refer the reader to~\cite{allen_multiscale_2017} for a detailed description of the axioms and the properties of the CP).

Shannon’s information theory provides a foundational framework for defining such functions. Following the approach in~\cite{gheorghiu-svirschevski_multiscale_2004}, which applied the CP to the infinite-range Ising model, we adopt the Shannon entropy as an information function. This is motivated by the probabilistic nature of the model and the correspondence between the Shannon entropy in information theory and the Gibbs entropy in statistical mechanics~\cite{jaynes_information_1957}. Its interpretation depends on the context in which it is applied: it quantifies either the uncertainty associated with a subsystem before measuring its state or, equivalently, the information gained as a result of the measurement~\cite{nielsen_quantum_2010}.

Let $\sigma^{(u)}=(\sigma_1^{(u)}, \ldots,  \sigma_{N-j}^{(u)})$ denote a spin configuration of the $u$th subsystem $\boldsymbol{\sigma}^{(u)}$, where the parentheses indicate a tuple of spin values, in contrast to the braces $\{\cdot\}$ used earlier to denote a subsystem of components. Given a probability distribution $p(\sigma^{(u)})$ that implicitly captures the interdependence among the $N-j$ spins, the corresponding Shannon entropy is defined as follows~\cite{cover_elements_2005}:
\begin{equation}\label{Equation: ShannonEntropy}
    H(\sigma_1^{(u)}, \ldots ,\sigma_{N-j}^{(u)}) = -\sum_{\sigma^{(u)}} p(\sigma^{(u)})\log_2{p(\sigma^{(u)})},
\end{equation}
where the sum is taken over all possible joint configurations of the subsystem, and the logarithm base 2 indicates that the entropy is measured in units of bits. 

With the Shannon entropy as the information function that describes subsystem $\boldsymbol{\sigma}^{(u)}$, the strength of the dependency among its spins is quantified by the multivariate conditional mutual information (CMI) $I(\cdot|\cdot)$. This measure is derived from the marginal Shannon entropies via the inclusion-exclusion principle and is given by~\cite{matsuda_physical_2000}
\begin{widetext}
\begin{equation}\label{Equation: ConditionalMutualInformation}
I(\sigma_1^{(u)}; \ldots; \sigma_{k}^{(u)} | \sigma_1^{(u^{\prime})}, \ldots, \sigma_{N-k}^{(u^{\prime})}) = \sum_{j=1}^k (-1)^{j+1}\sum_{\boldsymbol{\sigma}^{(v)}} H(\sigma_1^{(v)}, \ldots, \sigma_{j}^{(v)}|\sigma_1^{(u^{{\prime}})}, \ldots, \sigma_{N-k}^{(u^{\prime})}),
\end{equation}
%
where the inner summation is over all possible subsets $\boldsymbol{\sigma}^{(v)}\subseteq\{\sigma_1^{(u)}, \ldots, \sigma_{k}^{(u)}\}$ of size $j$, and the conditional Shannon entropy $H(\cdot|\cdot)$ is given by~\cite{cover_elements_2005}
%
\begin{equation}\label{Equation: ConditionalEntropyMarginal}
    H(\sigma_1^{(v)}, \ldots, \sigma_{j}^{(v)}|\sigma_1^{(u^{{\prime}})}, \ldots, \sigma_{N-k}^{(u^{\prime})}) = H(\sigma_1^{(v)}, \ldots, \sigma_j^{(v)}, \sigma_1^{(u^{\prime})}, \dots, \sigma_{N-k}^{(u^{\prime})}) - H(\sigma_1^{(u^{{\prime}})}, \ldots, \sigma_{N-k}^{(u^{\prime})}).
\end{equation}
\end{widetext}
Here, $\boldsymbol{\sigma}^{(u^{\prime})}=\{\sigma_1^{(u^{\prime})}, \ldots, \sigma_{N-k}^{(u^{\prime})}\}$ denotes the complement of subsystem $\boldsymbol{\sigma}^{(u)}$, comprising the remaining $N-k$ spins in system $\boldsymbol{\sigma}$. The CMI thus quantifies the amount of information shared among the spins in $\boldsymbol{\sigma}^{(u)}$, given that the configuration of the rest of the system $\sigma^{(u^{\prime})}$ is known~\cite{cover_elements_2005}. Note that the notations $\sigma_1^{(u)}, \ldots, \sigma_{k}^{(u)}$ and $\sigma_1^{(u)}; \ldots; \sigma_{k}^{(u)}$ respectively denote a joint subsystem and a mutual dependency. Both the Shannon entropy and the (conditional) mutual information are symmetric under any permutation of the involved spins.

By expanding Eq.~\eqref{Equation: ComplexityProfile} and using Eq.~\eqref{Equation: ConditionalMutualInformation}, the CP can be reformulated in terms of the CMI among $k$ \textit{or more} spins as
\begin{equation}\label{Equation: ComplexityProfileReExpressed}
    C(k) = \sum_{k^{\prime}=k}^N  \sum_{\boldsymbol{\sigma}^{(u)}} I(\sigma_1^{(u)}; \ldots; \sigma_{k^{\prime}}^{(u)} | \sigma_1^{(u^{\prime})}, \ldots, \sigma_{N-k^{\prime}}^{(u^{\prime})}).
\end{equation}
This form of the CP is particularly useful for computing the scale-specific shared information $D(k)$
\begin{gather}\label{Equation: ScaleSpecificInformation}
    \begin{alignedat}{1}
        D(k) &= C(k) - C(k+1)\\[5pt]
             &= \sum_{\boldsymbol{\sigma}^{(u)}} I(\sigma_1^{(u)}; \ldots; \sigma_{k}^{(u)} | \sigma_1^{(u^{\prime})}, \ldots, \sigma_{N-k}^{(u^{\prime})}),
    \end{alignedat}
\end{gather}
which characterizes the amount of information that is \textit{exclusively} shared among exactly $k$ spins. The computational complexity of evaluating Eqs.~\eqref{Equation: ComplexityProfileReExpressed} and~\eqref{Equation: ScaleSpecificInformation} depends on the number of spin subsystems, as indicated by the summation term, and/or symmetries of the system at scale $k$. Since spins in symmetry-related subsystems have the same shared information, these equations can be reformulated as a sum over distinct information terms, each weighted by its multiplicity factor. Therefore, systems with a higher symmetry require fewer terms in the computation of $C(k)$ and $D(k)$.

\section{Ising model: a paradigm for complex systems}\label{Section: IsingModel}
In the previous section, we defined the CP for the Ising model in terms of subsystem Shannon entropies and CMI. We now review several key concepts of the model that are essential for constructing probability distributions over spin subsystems, which, in turn, are required to apply these information-theoretic measures. For the remainder of this paper, we denote the $i$th spin $s_i$ in the $u$th subsystem as $s_j^{(u)}$, where $j$ is the index of the spin within that subsystem. We use $s_j^{(u)}$ and $\sigma_j^{(u)}$ when referring to the corresponding spin in the full system $\boldsymbol{s}$ and its subsystem $\boldsymbol{\sigma}^{(u)}$, respectively. When the superscript $u$ is omitted, the notation refers to the spin in the system as a whole.

\subsection{Statistical mechanics and critical behavior}\label{Subsection: StatisticalMechanicsAndCriticalBehavior}
For a system with $N$ Ising spins $\boldsymbol{s}=\{s_1, s_2, ..., s_N\}$, $s_i\in\{+1, -1\}$, we consider the following zero-field Hamiltonian:
\begin{equation}\label{Equation: IsingHamiltonian}
    \mathcal{H}(s)= - J\sum_{\left<i, j\right>} s_i s_j,
\end{equation}
where $J =+1$ is the ferromagnetic spin-spin coupling constant, and the sum is over all distinct nearest-neighbor pairs. The equilibrium distribution over the state space of the system is characterized by the Gibbs-Boltzmann probability distribution
\begin{equation}\label{Equation: BoltzmannDistribution}
    p(s) = \frac{1}{\mathcal{Z}}\exp\left[\beta J\sum_{\left<i, j\right>}s_is_j\right],
\end{equation}
where $\beta = \left(k_B T\right)^{-1}$ is the inverse temperature and $k_B$ is the Boltzmann constant that we assume to be equal to $1$ for simplicity. $\mathcal{Z}$ is the canonical partition function that encompasses all possible microstates of the system. 

\begin{figure*}[!ht]
\centering
\includegraphics[width=\textwidth,height=\textheight, keepaspectratio]{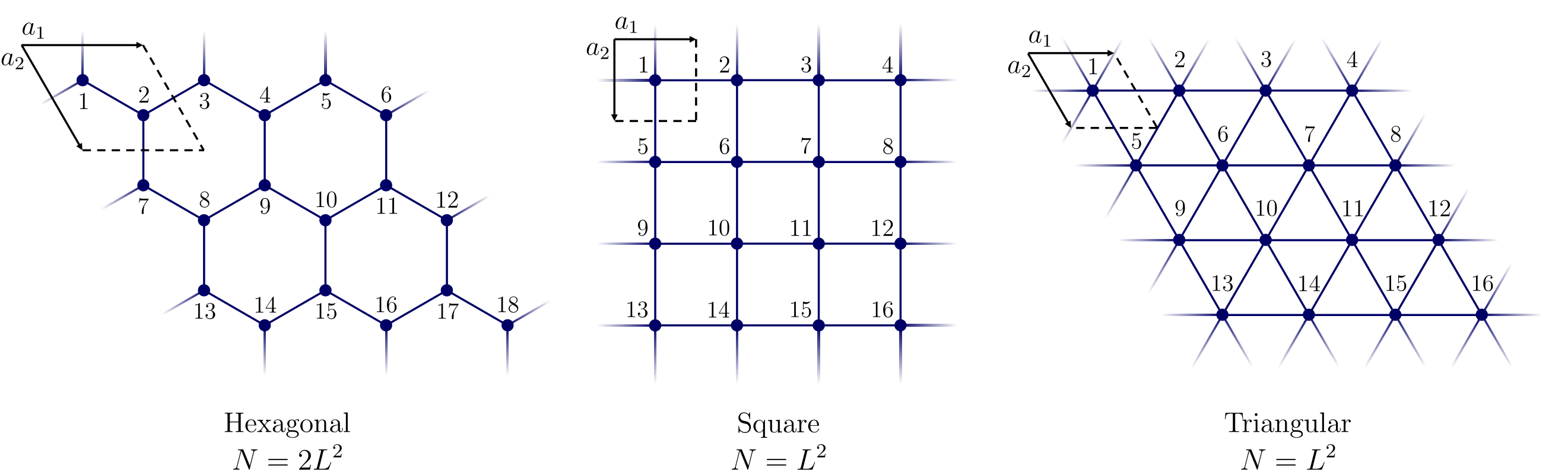}
\caption{The hexagonal, square and triangular lattices with periodic boundary conditions upon which we define the Ising model. The labeling of the spin sites is for the proper enumeration of the spin subsystems and correlation functions.}
\label{Figure: LatticeDefinitions}
\end{figure*}

As $\beta$ is increased, the system undergoes a second-order phase transition from a disordered phase to an ordered phase, in which all spins become mutually aligned. The transition occurs at a critical point $\beta_c$, whose value is characteristic of the considered lattice geometry and is analytically defined in the thermodynamic limit~\cite{baxter_exactly_2007}. In this work, we consider the hexagonal, square, and triangular lattices with a row-major labeling scheme, as illustrated in Fig.~\ref{Figure: LatticeDefinitions}. This is to study the impact of the number of nearest neighbors $z$ on the complexity of the system when the total number of spins is fixed. To reduce finite-size effects and preserve the translational symmetry of the model, we apply periodic boundary conditions. 

\subsection{Subsystem probability distribution}\label{Subsection: SubsystemProbabilityDistribution}
The probability distribution $p(\sigma^{(u)})$ for each subsystem $\boldsymbol{\sigma}^{(u)}$ can be defined in terms of the marginal probability distribution~\cite{cover_elements_2005}. Specifically, the probability of observing configuration $\sigma^{(u)}$ of a subsystem with $k$ spins is obtained by summing over the probabilities of all the $2^{N-k}$ configurations of the remaining $N-k$ spins in the system. As an illustrative example, consider a system with three spins $\boldsymbol{s} = \{s_1, s_2, s_3\}$. The probability of observing subsystem $\boldsymbol{\sigma}^{(u)}=\{\sigma_1^{(u)}, \sigma_2^{(u)}\}$ in configuration $\sigma^{(u)}=(\uparrow_1^{(u)}, \uparrow_2^{(u)})$ is given by
\begin{equation}\label{Equation: ExampleMarginalProbability}
    p(\uparrow_1^{(u)}, \uparrow_2^{(u)}) = p(\uparrow_1, \uparrow_2, \uparrow_3) + p(\uparrow_1, \uparrow_2, \downarrow_3),
\end{equation}
where the joint system probability $p(s)$ is defined by the Gibbs-Boltzmann distribution in Eq.~\eqref{Equation: BoltzmannDistribution}. For large systems, it is computationally convenient to express this marginal probability strictly as a function of the states of the spins of the subsystem. This is achieved by expressing this probability as a function of the subsystem configuration $\sigma^{(u)}$ and spin correlation functions $\langle s_1^{(v)}\cdots s_{\ell}^{(v)}\rangle_{\mathcal{H}}$ as follows (see Appendix~\ref{Appendix: MarginalProabilityDistribution}):
\begin{align}\label{Equation: MarginalProbabilityDistribution}
p(\sigma_{1}^{(u)}, \ldots, \sigma_k^{(u)}) &= \frac{1}{2^k}\bigg(1+\sum_{\ell \text{ even}}^k \sum_{\boldsymbol{\sigma}^{(v)}} \sigma^{(v)}_1\cdots\sigma^{(v)}_{\ell} \nonumber \\[5pt]
&\quad \times \langle s_1^{(v)}\cdots s_{\ell}^{(v)}\rangle_{\mathcal{H}}\bigg).
\end{align}
The inner summation is over all subsets $\boldsymbol{\sigma}^{(v)}\subseteq\{\sigma_1^{(u)}, \ldots, \sigma_{k}^{(u)}\}$ of size $\ell$, and $\left< \cdot \right>_{\mathcal{H}}$ denotes the thermal average with respect to the Hamiltonian $\mathcal{H}(s)$. In the absence of an external magnetic field, all odd-point correlation functions vanish due to the up and down symmetry of the Ising model, that is, the Hamiltonian is invariant under a global spin flip. 

This marginal probability is implicitly dependent on the full probability distribution in Eq.~\eqref{Equation: BoltzmannDistribution}, as the correlation functions encode the coupling between the subsystem and the rest of the system. We also note that for $k=N$, Eq.~\eqref{Equation: MarginalProbabilityDistribution} corresponds to a special case ($t=0$) of the joint probability distribution of the time-dependent Ising model, as described by Glauber dynamics in~\cite{glauber1963time}.

\subsection{Computational complexity}
\label{Subsection: ComputationalComplexity}
Since the CP in Eq.~\eqref{Equation: ComplexityProfile} does not have a simple closed-form expression, it rapidly becomes computationally intractable for large systems. Due to its inherent combinatorial form, it exhibits a time complexity of $\Omega(N!)$, where, at minimum, the computational time grows factorially with the size of the system. This complexity arises from the necessity to enumerate all possible subsystems when computing the total subsystem information $Q$ in Eq.~\eqref{Equation: SubsystemInformation}. To mitigate such complexity, two main approaches may be implemented: approximating the form of the CP and/or leveraging some properties of the considered phenomenon.

In~\cite{bar-yam_computationally_2013}, the CP was approximated using pairwise information functions. This approximation satisfies the main properties of the original formulation~\cite{allen_multiscale_2017} and allows for a polynomial-time computation. However, it does not capture higher-order information correlations that cannot be represented by dependencies between two components. For systems with strong emergence~\cite{bar-yam_mathematical_2004}, one of which is the Ising model, this approximation may limit the ability of the CP to fully capture correlations that are inherently characteristic of large-scale behaviors.

The other approach is to leverage specific properties of a system or apply mean-field approximations, as demonstrated for the infinite-range Ising model in~\cite{gheorghiu-svirschevski_multiscale_2004}. In this case, the permutational symmetry, at all scales, of both the exact and mean-field models allows the probability distribution to be expressed as a function of the total magnetization. As a result, all subsystems with $k$ spins have identical Shannon entropy, reducing the total number of informational quantities needed to compute $Q$ from $2^N$ to $N$. However, while a mean-field approximation can capture certain macroscopic features such as the existence of phase transitions, it typically reduces a complex system to an effective one-body problem, often neglecting correlations between components responsible for emergent large-scale behaviors.

In this paper, we make use of the Markov properties of the short-range Ising model to simplify the computation of subsystem Shannon entropies of large systems. As we will see in Sec.~\ref{Subsection: MarkovRandomFields}, these properties effectively reduce the number of DoF in the marginal probability distribution, therefore restricting the summation in Eq.~\eqref{Equation: ShannonEntropy}, which spans an exponentially large number of configurations, to a smaller number of configurations. Note that this is not an approximation of the model, but an intrinsic property of finite-range Ising systems. Additionally, we exploit the spatial symmetries of the lattice geometries considered in Fig.~\ref{Figure: LatticeDefinitions} to express $Q$ in terms of distinct subsystem information functions weighted by their multiplicity factors.

\subsection{Markov random fields}\label{Subsection: MarkovRandomFields}
A Markov field is a model for representing high-dimensional probability distributions by mapping conditional dependencies among random variables onto undirected graphs~\cite{hamilton_information_2017}. The Ising model is a special case of this generalized framework, in which local interactions between neighboring spins are encoded by an undirected graph representing the lattice geometry. This graphical representation is particularly useful, as it enables us to leverage the properties of conditional independence to infer the distributions of spin subsystems.

\begin{figure}[!ht]
\centering
\includegraphics[width=0.495\textwidth,height=\textheight, keepaspectratio]{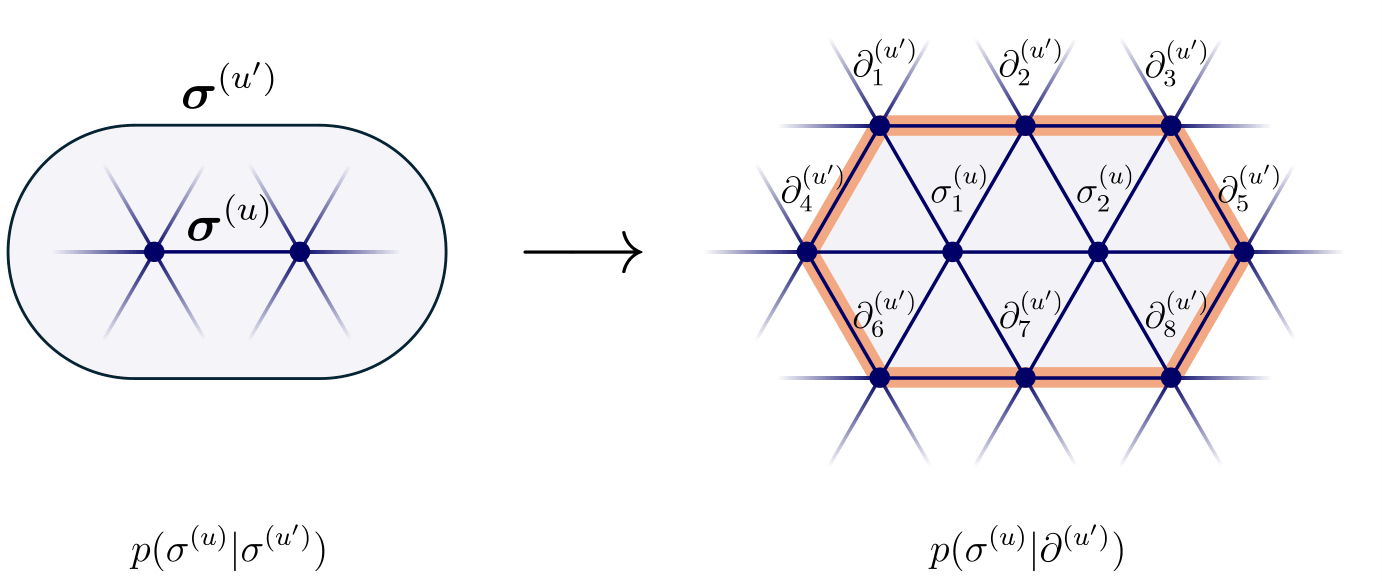}
\caption{The reduction in the number of DoF in the conditional probability $p(\sigma^{(u)}|\sigma^{(u^{\prime})})$ using the Markov field property of conditional independence. The highlighted spin subsystem $\boldsymbol{\partial}^{(u^{\prime})}$ is a subset of $\boldsymbol{\sigma}^{(u^{\prime})}$ and is the boundary that separates subsystem $\boldsymbol{\sigma}^{(u)}$ from the rest of the system.}
\label{Figure: MarkovBlanket}
\end{figure}

The Hammersley-Clifford theorem~\cite{clifford1971markov} states that all classical Gibbs distributions with finite-range interactions, including the distribution in Eq.~\eqref{Equation: BoltzmannDistribution}, are Markov fields and satisfy the Markov property of conditional independence. For the Ising model, this implies that the probability of observing the configuration of subsystem $\boldsymbol{\sigma}^{(u)}$ is conditionally independent of the rest of the system $\boldsymbol{\sigma}^{(u^{\prime})}$ once the configuration of the \textit{Markov blanket} or \textit{boundary} $\partial^{(u^{\prime})}$ separating them is known:
\begin{equation}\label{Equation: ConditionalIndependence}
    p(\sigma^{(u)}|\sigma^{(u^{\prime})}) = p(\sigma^{(u)}|\partial^{(u^{\prime})}) \quad \text{where} \quad \boldsymbol{\partial}^{(u^{\prime})}\subset \boldsymbol{\sigma}^{(u^{\prime})}.
\end{equation}
The boundary $\boldsymbol{\partial}^{(u^{\prime})}$ is defined as the subset of spins that intersects every direct path from any spin in $\boldsymbol{\sigma}^{(u)}$ to any spin in $\boldsymbol{\sigma}^{(u^{\prime})} \setminus \boldsymbol{\partial}^{(u^{\prime})}$. In Fig.~\ref{Figure: MarkovBlanket}, for example, the subsystem $\boldsymbol{\sigma}^{(u)}=\{\sigma_1^{(u)}, \sigma_2^{(u)}\}$ contains two neighboring spins, and its separating boundary is formed by the highlighted eight spins in $\boldsymbol{\partial}^{(u^{\prime})}$. Note that while correlations between $\boldsymbol{\sigma}^{(u)}$ and $\boldsymbol{\sigma}^{(u^{\prime})} \setminus \boldsymbol{\partial}^{(u^{\prime})}$ may still exist, they are all mediated by the boundary. 

Given the Markov field property, we can alternatively compute the Shannon entropy in Eq.~\eqref{Equation: ShannonEntropy} for large subsystems using the joint system entropy $H(\boldsymbol{s})$ and the reduced conditional entropy, as follows:
\begin{gather}\label{Equation: JointEntropy}
    \begin{alignedat}{1}
        H(\boldsymbol{\sigma}^{(u^{\prime})}) &= H(\boldsymbol{\sigma}^{(u)}, \boldsymbol{\sigma}^{(u^{\prime})}) - H(\boldsymbol{\sigma}^{(u)}|\boldsymbol{\sigma}^{(u^{\prime})})\\[5pt]
                                 &= H(\boldsymbol{s}) - H(\boldsymbol{\sigma}^{(u)}|\boldsymbol{\partial}^{(u^{\prime})}).\\[5pt]
    \end{alignedat}
\end{gather}
The conditional entropy, given in Eq.~\eqref{Equation: ConditionalEntropyMarginal}, is expressed in terms of the joint and conditional probability distributions as~\cite{cover_elements_2005}:
\begin{align}\label{Equation: ConditionalEntropyProbability}
H(\boldsymbol{\sigma}^{(u)}|\boldsymbol{\partial}^{(u^{\prime})}) &= -\sum_{\partial^{(u^{\prime})}} p(\partial^{(u^{\prime})}) \sum_{\sigma^{(u)}} p(\sigma^{(u)}|\partial ^{(u^{\prime})}) \nonumber \\[5pt]
&\quad \times \log_2 p(\sigma^{(u)}|\partial ^{(u^{\prime})}).
\end{align}
The probability of the configuration of the boundary $p(\partial^{(u^{\prime})})$ is given by Eq.~\eqref{Equation: MarginalProbabilityDistribution}. Following the definition of the conditional probability $p(\sigma^{(u)}|\partial ^{(u^{\prime})})$~\cite{cover_elements_2005}, we derive a generalized expression in Appendix~\ref{Appendix: ConditionalProbability} that is independent of the system size for a given geometry of the boundary $\boldsymbol{\partial}^{(u^{\prime})}$. 

\subsection{Computational details}\label{Subsection: ComputationalDetails}
The complexity at all scales for a small number of spins (up to $N\sim 20$) can be computed exactly, using the form of the marginal probability given in Eq.~\eqref{Equation: ExampleMarginalProbability}. For a large number of spins, evaluating the subsystem Shannon entropy becomes more computationally tractable using the expression in Eq.~\eqref{Equation: JointEntropy}, which requires two quantities: the joint system entropy and the reduced conditional entropy. To estimate $H(\boldsymbol{s})$ and the correlation functions $\langle s_1^{(v)}\cdots s_{\ell}^{(v)}\rangle_{\mathcal{H}}$ needed for the probability distribution $p(\partial^{(u^{\prime})})$, we implement Monte Carlo simulations of the Ising model.

The Shannon entropy is related to the Gibbs entropy $S$ through the relation
\begin{equation}\label{Equation: GibbsEntropy}
    H(\boldsymbol{s}) = \frac{S}{k_B\ln 2},
\end{equation}
which requires fine resolution in $\beta$ to obtain accurate estimates through numerical integration of the specific heat~\cite{newman1999monte}. This becomes particularly challenging near the phase transition, where thermodynamic quantities of large systems exhibit singularlike behavior. Since $S$ cannot be directly estimated from standard Monte Carlo simulations, we instead compute it from the density of states of the system, estimated using the replica-exchange Wang-Landau algorithm (see Appendix~\ref{Appendix: REWL}). This method allows us to estimate $S$, and therefore $H(\boldsymbol{s})$, at any finite temperature. 

For a given boundary geometry $\boldsymbol{\partial}^{(u^{\prime})}$, the exact form of the reduced conditional probability $p(\sigma^{(u)}|\partial ^{(u^{\prime})})$ is provided in Appendix~\ref{Appendix: ConditionalProbability}. The joint probability $p(\partial^{(u^{\prime})})$ is expressed in terms of the even-point correlation functions of all possible combinations of the boundary spins. These correlations can be estimated directly using a standard Monte Carlo simulation of the Ising model, as follows:
\begin{gather}\label{Equation: CorrelationFunctionEstimate}
\begin{alignedat}{1}
    \langle s_1^{(v)}\cdots s_{\ell}^{(v)}\rangle_{\mathcal{H}} &= \frac{1}{\mathcal{Z}}\sum_s s_1^{(v)}\cdots s_{\ell}^{(v)} e^{-\beta \mathcal{H}(s)}\\[5pt]
                                                &\approx \frac{1}{M}\sum_{i=1}^M q_{\mu_i},
\end{alignedat}
\end{gather}
where the first expression denotes the ensemble average over all system configurations, and the second is its approximation using $M$ statistically independent samples at thermal equilibrium. The quantity $q_{\mu_i}\in\{+1, -1\}$ is the instantaneous value of the correlation function in microstate $\mu_{i}$. To efficiently obtain a large number of such measurements, we implement a $64$-bit multispin-coded Metropolis algorithm that enables multiple independent simulations in parallel using a single processor (see Appendix~\ref{Appendix: MSC}). Note that we only estimate the nonequivalent correlation functions, as many are identical across different subsystems due to the symmetries of the considered lattice geometries. We identify these correlation subsets using the point group symmetries of each lattice, that is, translational, rotational, and inversion symmetries. Further details about the computational implementation and simulation setup are provided in Appendix~\ref{Appendix: MCSimulation}, and the corresponding simulation scripts are available on Zenodo~\cite{al_azki_2025_15515297}.

\section{Complexity profile analysis}\label{Section: ComplexityProfileAnalysis}
In Secs. \ref{Subsection: SubsystemProbabilityDistribution} and \ref{Subsection: MarkovRandomFields}, we introduced the subsystem probability distribution that is explicitly defined in terms of the configuration of the subsystem, and demonstrated how the number of DoF in this probability can be reduced using the Markov properties of conditional independence. Nevertheless, given that the computational complexity of the CP scales factorially with the number of spins, we limited its exact computation at all scales to small system sizes for the three lattices. We validated our results by comparing the Shannon entropies computed using the two forms of the subsystem probabilities in Eqs.~\eqref{Equation: ExampleMarginalProbability} and~\eqref{Equation: MarginalProbabilityDistribution}. While such systems do not exhibit true phase transitions due to finite-size effects, the CPs nonetheless capture key features of the phase transition observed in the Ising model in the thermodynamic limit. Additionally, to gain insight into the behavior of higher-order information correlations near the transition point, we estimated the complexity and scale-specific shared information at scale $k=2$ for larger systems using Monte Carlo simulations (see Sec.~\ref{Subsection: PairwiseComplexity}).

\subsection{Full-scale complexity}\label{Subsection: FullScaleComplexity}
\begin{figure*}[!ht]
\centering
\begin{subfigure}{0.496\textwidth}
    \begin{tikzpicture}
    \node[] (pic) at (0,0) {\includegraphics[width=\textwidth]{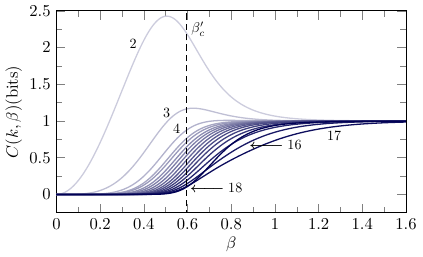}};
    \end{tikzpicture}
    \caption{Hexagonal lattice}
\end{subfigure}
\hfill
\begin{subfigure}{0.496\textwidth}
\begin{tikzpicture}
    \node[] (pic) at (0,0) {\includegraphics[width=\textwidth]{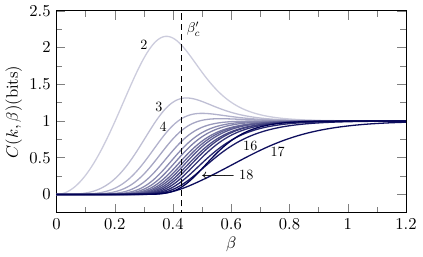}};
    \end{tikzpicture}
    \caption{Square lattice}
\end{subfigure}
\medskip
\begin{subfigure}{0.496\textwidth}
\begin{tikzpicture}
    \node[] (pic) at (0,0) {\includegraphics[width=\textwidth]{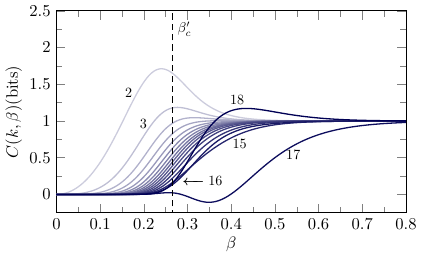}};
    \end{tikzpicture}
    \caption{Triangular lattice}
\end{subfigure}
\hfill
\begin{subfigure}{0.496\textwidth}
\begin{tikzpicture}
    \node[] (pic) at (0,0) {\includegraphics[width=\textwidth]{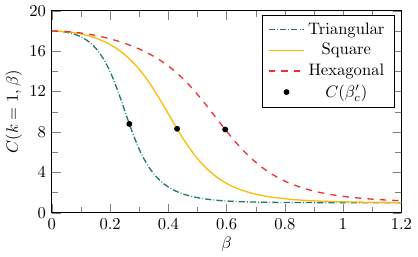}};
    \end{tikzpicture}
    \caption{Fine-scale complexity}
\end{subfigure}
\caption{The complexities as functions of $\beta$ for the ferromagnetic (a) hexagonal, (b) square and (c) triangular lattices with $N=18$ spins. The curves are labeled by scale from $k=2$ (light) to $k=18$ (dark). (d) The finest-scale complexity $C(1)$ for the three lattices, which corresponds to the joint Shannon entropy of the whole system $H(\boldsymbol{s})$.}
\label{Figure: LatticeComplexity}
\end{figure*}

We computed the CP at all scales exactly using Eq.~\eqref{Equation: ComplexityProfile} for the hexagonal, square, and triangular ferromagnetic lattices (see Fig.~\ref{Figure: LatticeDefinitions}), with $N=18$ spins, as functions of the inverse temperature $\beta$. The dashed lines in Figs.~\ref{Figure: LatticeComplexity}(a)–~\ref{Figure: LatticeComplexity}(c) indicate the pseudocritical temperatures $\beta_{c}^{\prime}$ of the finite-size lattices to mark the critical region, which we obtained from the position of the maximum of the specific heat of each lattice.

As we can see from Figs.~\ref{Figure: LatticeComplexity}(a)–~\ref{Figure: LatticeComplexity}(c), $C(k) \rightarrow 0$ as $\beta \rightarrow 0$ indicates that the disordered phase of the three systems is characterized by the absence of information correlations at scales $k\geq 2$. In contrast, the complexity curves at all scales converge asymptotically to unity in the ferromagnetic phase for large $\beta$. In this regime, since all spins are completely dependent, a single bit represents the minimum amount of information needed to describe the states of $k$ (or more) correlated spins. 

In Fig.~\ref{Figure: LatticeComplexity}(d), where we show the behavior of $C(k=1)$ for the three lattices, we see that the systems exhibit the emergence of microscopic structures towards the strongly disordered phase, as evidenced by the rapid increase of $C(k=1)$ near $\beta_{c}^{\prime}$, reaching its maximum at $\beta = 0$. We note that $C(k=1) = H(\boldsymbol{s})$, which corresponds to the joint entropy of the system. This suggests that the systems require $N$ independent bits of information (or spins) to fully describe their microstate, characteristic of systems with noninteracting spins. Therefore, in the context of the multiscale complexity formalism, the Ising model behaves as a simple system in both the high- and low-temperature regimes, with its complex behavior emerging in the critical region. 

\subsubsection{Scale-dependent complexity}
To understand the behavior of $C(k)$ in the critical region, we categorize the curves in Figs.~\ref{Figure: LatticeComplexity}(a)–~\ref{Figure: LatticeComplexity}(c) into fine-, intermediate-, and large-scale complexities, based on the number of correlated spins $k$. Fine-scale complexities ($k/N \ll 1$) exhibit decaying maxima as a function of scale near $\beta_c^{\prime}$, with the complexity at scale $k=2$ having the largest amplitude. This behavior is virtually absent in the infinite-range Ising model for the same system size (see Figs. 1(f) and 1(l) in~\cite{gheorghiu-svirschevski_multiscale_2004}). As we will see in Sec. \ref{Subsection: PairwiseComplexity}, this peak is likely an intrinsic information-theoretic characteristic of the disordered phase of classical spin systems. For intermediate scales ($0< k/N< 1$), the complexities generally exhibit an increasingly monotonic behavior with increasing $\beta$. However, at each temperature, larger scales exhibit reduced complexity, as the subset of dependencies contributing to $C(k)$ at larger scales is a subset of that at finer scales [see Eq.~\eqref{Equation: ComplexityProfileReExpressed}].
 
Large-scale complexities ($k/N\approx 1$), as represented by the dark curves, generally develop above $\beta_{c}^{\prime}$ and deviate from the monotonic decrease with increasing scale. For the triangular lattice, the complexity at scale $k=N-1$ reaches a negative minimum before asymptotically converging to unity. We observed a similar behavior for the square lattice at this scale from $N=20$ (see Appendix~\ref{Appendix: FullScaleComplexity}), and expect the same for the hexagonal lattice, although at larger system sizes. A similar behavior appears in the mutual information of frustrated spin systems ($J=-1$)~\cite{matsuda_physical_2000,matsuda_information_2001}. There, emergent correlations due to geometric frustration produce a global stabilizing effect that competes with the local anti-alignment of nearest neighbors near the temperature at which the negative minimum occurs.

\subsubsection{Negative complexity}
The physical interpretation of the negative values of the CP depends on the phenomenon exhibited by the system and can be linked to the origin of negative (conditional) mutual information. As suggested in~\cite{bar-yam_multiscale_2004}, negative mutual information results from conflicting inferences about the state of a system due to partial or incomplete knowledge of the states of its subsystems. This reflects the presence of intrinsic many-body correlations that cannot be represented by lower-order ones.  

For the ferromagnetic Ising model, the negative complexity arises from the competing effects of thermal fluctuations and the long-range order associated with the ensemble of magnetized states for $\beta > \beta_c^{\prime}$ (see Appendix~\ref{Appendix: FullScaleComplexity}). In the temperature regime where $C(k=N-1) < 0$, the system explores metastable ordered states, in which at least $N-1$ spins are mutually aligned along any of the two distinct macroscopic orientations. This contrasts with the behavior in the thermodynamic limit, where the up or down symmetry is spontaneously broken at $\beta_c$, and the system remains confined to a single symmetry-broken phase. Indeed, under a spin-flip dynamics simulation, thermal fluctuations result in a nonzero probability of transitioning between positively and negatively magnetized states~\cite{Faulkner2025}, with an occupation time that scales as $\exp{(\sqrt{N})}$~\cite{lebowitz_statistical_1999,fraser_spontaneous_2016}. The conflicting information about the state of the system, manifested as negative complexity, thus arises from the destruction of long-range order during the transition, as the system passes through intermediate, unmagnetized states. This behavior is counterintuitive since the system is expected to mostly occupy magnetized states within the temperature regime associated with long-range order.

For the triangular lattice, the system eventually transitions into one of the two symmetry-breaking ground states above $\beta\approx 0.4$, as indicated by the positive values and peak of the complexities at scales $k=N-1$ and $N$, respectively. This transition results from the dominance of long-range order over thermal fluctuations, which vanish for large $\beta$ (see Appendix~\ref{Appendix: FullScaleComplexity}). 

\subsubsection{Scale-specific shared information}
\begin{figure*}[!ht]
\centering
\begin{subfigure}{0.496\textwidth}
    \begin{tikzpicture}
    \node[] (pic) at (0,0) {\includegraphics[width=\textwidth]{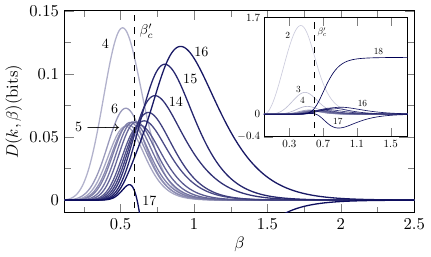}};
    \end{tikzpicture}
    \caption{Hexagonal lattice}
\end{subfigure}
\hfill
\begin{subfigure}{0.496\textwidth}
\begin{tikzpicture}
    \node[] (pic) at (0,0) {\includegraphics[width=\textwidth]{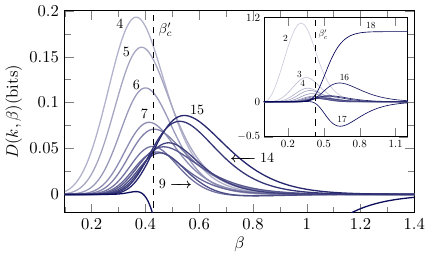}};
    \end{tikzpicture}
    \caption{Square lattice}
\end{subfigure}
\medskip
\begin{subfigure}{0.496\textwidth}
\begin{tikzpicture}
    \node[] (pic) at (0,0) {\includegraphics[width=\textwidth]{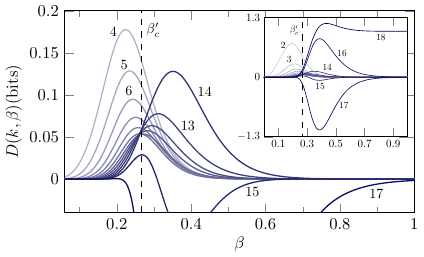}};
    \end{tikzpicture}
    \caption{Triangular lattice}
\end{subfigure}
\hfill
\begin{subfigure}{0.496\textwidth}
\begin{tikzpicture}
    \node[] (pic) at (0,0) {\includegraphics[width=\textwidth]{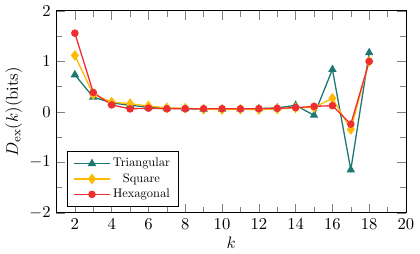}};
    \end{tikzpicture}
    \caption{Extremal values of $D(k, \beta)$}
\end{subfigure}
\caption{(a) The scale-specific shared information for the ferromagnetic (a) hexagonal, (b) square and (c) triangular lattices with $N=18$ spins (the insets show the global behavior at scales $k\geq2$). (d) A comparison between the behavior of the extrema of $D(k)$ of each lattice as a function of scale. Note that these extrema correspond to different $\beta$ as shown in (a)-(c).}
\label{Figure: ScaleSpecificComplexity}
\end{figure*}

To identify the relevant scales at each temperature, we also computed the scale-specific shared information $D(k)$ as a function of $\beta$ for each lattice [Figs.~\ref{Figure: ScaleSpecificComplexity}(a)–~\ref{Figure: ScaleSpecificComplexity}(c)], using Eq.~\eqref{Equation: ScaleSpecificInformation}. As a reminder, $D(k)$ quantifies the information that is \textit{exclusively} shared among $k$ spins. The extrema of $D(k)$ mark the characteristic temperatures at which subsystems with $k$ spins contribute maximally to the shared information among the spins at scale $k$ or larger. High- and low-temperature regimes are dominated by fine and large scales, respectively, as indicated by the positions of the extrema, while $D(k)$ at intermediate scales exhibits small-amplitude maxima near $\beta_c^{\prime}$. From the insets, for $k\geq 2$, we observe that $D(k=2)$ dominates over other scales for $\beta <\beta_c$ and persists slightly beyond $\beta_c$.

The difference in the behavior of $D(k)$ across different scales, as well as between subcritical and supercritical regimes can be clearly illustrated by plotting its extremal values, $D_{\text{ex}}(k)$, against scale, as shown in Fig.~\ref{Figure: ScaleSpecificComplexity}(d). From Figs.~\ref{Figure: ScaleSpecificComplexity}(a)–~\ref{Figure: ScaleSpecificComplexity}(c), for $k<9$, $D(k)$ peaks below $\beta_{c}^{\prime}$, whereas for $k\geq9$, the extremal values occur above $\beta_{c}^{\prime}$. For all three systems, the absolute value of $D_{\text{ex}}(k)$ decreases monotonically with increasing scale up to $k=12$, beyond which it increases again, exhibiting an oscillatory behavior. 

At fine scales ($k\leq 4$), the decay of $D_{\text{ex}}(k)$ for $\beta < \beta_c^{\prime}$ reflects the thermal fluctuations on short length scales, where spins are only correlated over short distances. At intermediate scales ($5\leq k\leq 14$), $D_{\text{ex}}(k)$ remains small and is mainly centered around $\beta_c^{\prime}$. This may indicate the emergence of scale invariance, a signature of criticality, where the correlation length approaches the linear size of the system. However, confirming this behavior requires computing $D(k)$ for larger systems. The temperature dependence of the peak positions, along with the strong overlap of the curves near $\beta_c^{\prime}$ in Figs.~\ref{Figure: ScaleSpecificComplexity}(a)–~\ref{Figure: ScaleSpecificComplexity}(c), suggest the formation of magnetic domains and changes in their size distribution as the temperature decreases (see Sec. \ref{Section: Discussion}).

At scales comparable to the size of the system ($k\geq 15$), $D_{\text{ex}}(k)$ exhibits oscillatory behavior with amplitudes that increase with scale. Generally, such behavior has been associated with the emergence of collective phenomena arising from global or local constraints on the behavior of individual components~\cite{bar-yam_mathematical_2004}. Here, it reflects the emergence of an order parameter that describes the existence of an ordered phase. However, the amplitude of these oscillations is significantly smaller than that observed in the infinite-range model~\cite{gheorghiu-svirschevski_multiscale_2004}. This suggests, as expected, that in lattice systems, the ferromagnetic ordering constraint, where a macroscopic magnetization results from the mutual alignment of spins, is enforced locally through nearest neighbors. In contrast, in the infinite-range model, where each spin interacts with the rest of the system, the alignment constraint is enforced globally across the entire system.

\subsection{Complexity versus scale tradeoff}\label{Subsection: ComplexityVsScale}
\begin{figure}[!ht]
\centering
\includegraphics[width=0.325\textwidth,height=\textheight, keepaspectratio]{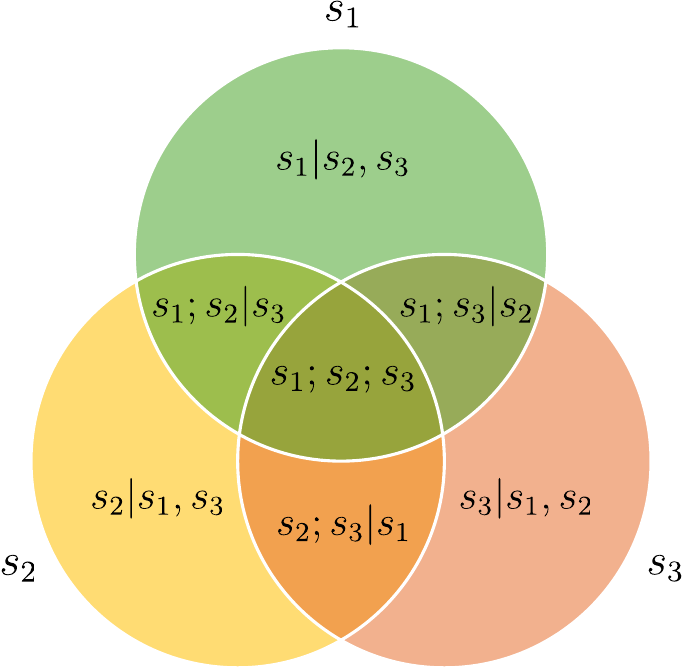}
\caption{Venn diagram representing the dependency space of spin system $\boldsymbol{s}=\{s_1, s_2, s_3\}$.}
\label{Figure: VennDiagram}
\end{figure}

The structure of a system can be represented diagrammatically using high-dimensional Venn diagrams (see Fig.~\ref{Figure: VennDiagram}). This representation depicts the dependency space of the system, which is determined collectively by all possible dependencies. In our case, while each dependency is characterized by the CMI, their significance within this space is determined by their scale, that is, the number of spins involved. 

For a system with a fixed number of components, such as the Ising model, the total scale-weighted CMI contained within its dependency space is a conserved quantity~\cite{allen_multiscale_2017,siegenfeld_introduction_2020}, and the CP therefore satisfies the following sum rule:
\begin{equation}\label{ComplexityConservation}
    \sum_{k=1}^N C(k) = N \text{ bits}.
\end{equation}
This implies that for an arbitrary Ising system with $N$ microscopic DoF, the sum of the CP across all scales remains constant for any given temperature. However, the behaviors of $C(k)$ and $D(k)$ vary quantitatively among different lattices, particularly at fine and large scales (see Figs.~\ref{Figure: LatticeComplexity} and~\ref{Figure: ScaleSpecificComplexity}). This variation, constrained by the sum rule, reflects a tradeoff between complexity and scale, which arises from the redistribution of shared information and the arrangement of spins in each lattice.

\begin{figure*}[!ht]
\centering
\begin{subfigure}{0.496\textwidth}
    \begin{tikzpicture}
    \node[] (pic) at (0,0) {\includegraphics[width=\textwidth]{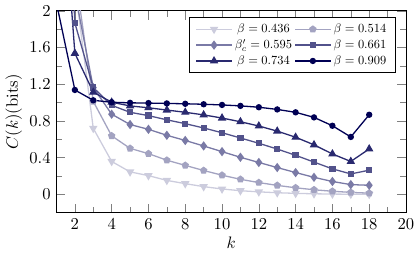}};
    \end{tikzpicture}
    \caption{Hexagonal lattice}
\end{subfigure}
\hfill
\begin{subfigure}{0.496\textwidth}
\begin{tikzpicture}
    \node[] (pic) at (0,0) {\includegraphics[width=\textwidth]{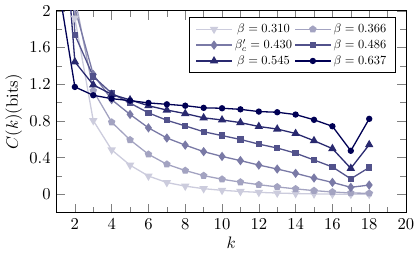}};
    \end{tikzpicture}
    \caption{Square lattice}
\end{subfigure}
\medskip
\begin{subfigure}{0.496\textwidth}
\begin{tikzpicture}
    \node[] (pic) at (0,0) {\includegraphics[width=\textwidth]{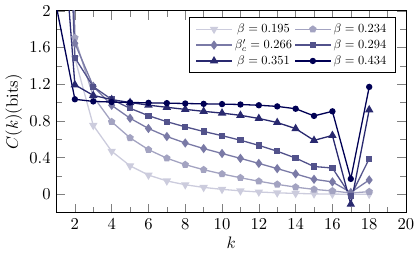}};
    \end{tikzpicture}
    \caption{Triangular lattice}
\end{subfigure}
\hfill
\begin{subfigure}{0.496\textwidth}
\begin{tikzpicture}
    \node[] (pic) at (0,0) {\includegraphics[width=\textwidth]{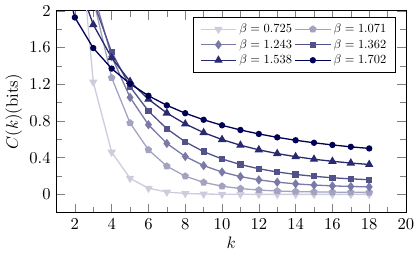}};
    \end{tikzpicture}
    \caption{1D Ising model}
\end{subfigure}
\caption{The complexity profiles as functions of scale $k$ for the (a) hexagonal, (b) square, and (c) triangular lattices, and (d) the one-dimensional Ising model with $N=18$ spins. Each profile represents a different temperature, ranging from high (light) to low (dark) temperatures.}
\label{Figure: ComplexityScaleTradeoff}
\end{figure*}

To illustrate this tradeoff, we plot in Figs.~\ref{Figure: ComplexityScaleTradeoff}(a)–~\ref{Figure: ComplexityScaleTradeoff}(c) the CP against scale $k$ for the three lattices at different temperatures. As discussed in the previous section, the non-monotonic behavior at large scales signifies the emergence of macroscopic magnetization, which occurs for all three lattices near $\beta\approx\beta_c^{\prime}$. Notably, the area under each profile remains constant ($N$ bits), thereby satisfying the sum rule in Eq.~\eqref{ComplexityConservation} and illustrating the tradeoff between complexity and scale. 

To demonstrate the universality of the sum rule in systems with identical microscopic DoF, Fig.~\ref{Figure: ComplexityScaleTradeoff}(d) shows the CPs for a 1D Ising model with $N=18$ spins and nearest-neighbor ferromagnetic interactions. In contrast to the 2D model, this model does not exhibit a finite-temperature phase transition. The profiles decrease monotonically with increasing scale at all temperatures, with greater complexity observed at larger scales as the temperature decreases. Note that as $\beta \rightarrow \infty$, the absence of thermal fluctuations leads to the mutual alignment of spins, resulting in $C(k)=1$ bit for all $k$. 

Due to the conservation of the sum of the CP at all scales, we notice in Fig.~\ref{Figure: ComplexityScaleTradeoff} that at high temperatures, the Ising model exhibits high complexity at fine scales, whereas at low temperatures, complexity decreases but extends to larger scales. This observation indicates that large-scale behaviors emerge from the coordination of multiple spins, whose behaviors are constrained by the dependencies among them. Consequently, the system exhibits a tradeoff between \textit{adaptability} and \textit{efficiency}, as defined in~\cite{siegenfeld_introduction_2020}, which is characteristic of complex systems with critical phenomena. Adaptability refers to the capacity of a system to respond rapidly to environmental fluctuations through independent behaviors of individual components, whereas efficiency arises from the coordinated operation of multiple components that enables the system to maintain macroscopic order by suppressing local perturbations.

In the strongly disordered phase, the high complexity at small scales allows the system to respond to thermal fluctuations or changes in temperature by flipping small clusters of correlated spins. This allows the system to explore a broad region of the state space to maintain its high-entropy state. At sufficiently low temperatures, the system becomes efficient at sustaining macroscopic magnetization, as collective constraints among spins make it increasingly resistant to such perturbations. In the critical region, all three lattices exhibit a balance between adaptability and efficiency, as evidenced by the emergence of the nonmonotonic behavior in the large-scale complexities at $\beta \approx \beta_c^{\prime}$. 

\subsection{Pairwise complexity}\label{Subsection: PairwiseComplexity}
We now examine the behavior of $C(k)$ and $D(k)$ for larger systems, where $L$, as illustrated in Fig.~\ref{Figure: LatticeDefinitions}, denotes the linear size of the system in lattice units. However, due to the computational complexity of both measures (see Sec. \ref{Subsection: ComputationalComplexity}), we restrict their computation to scales $k=2$ and $3$, where the latter is used to compute $D(k=2)$ using Eq.~\eqref{Equation: ScaleSpecificInformation} as follows:
\begin{equation}\label{Equation: DFromC}
D(k=2) = C(k=2) - C(k=3).
\end{equation}
Since the number of CMI terms contributing to $C(k=2)$ and $D(k=2)$ increases combinatorially with the size of the system, we consider the normalized quantities $c = C(k=2)/N$ and $d = D(k=2) / N$. As will see in Sec.~\ref{Section: Discussion}, the subsets contributing to these quantities are those that form connected clusters of spins. Enumerating such subsets is equivalent to the problem of enumerating lattice animals~\cite{PhysRevLettParisi}, for which no closed-form solution exists, even in 2D. Consequently, we choose $N$ as a normalization factor, and, as will be shown, both quantities scale linearly with $N$ in the disordered phase. It is important to note that this choice does not yield a per-site measure of complexity or shared information; rather, it illustrates the scaling behavior of these quantities in the thermodynamic limit.

\begin{figure*}[ht!]
\centering
\begin{subfigure}{0.328\textwidth}
    \begin{tikzpicture}
    \node[] (pic) at (0,0) {\includegraphics[width=\textwidth]{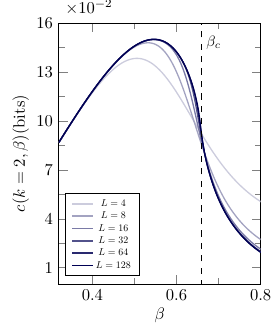}};
    \end{tikzpicture}
    \caption{Hexagonal lattice}
\end{subfigure}
\begin{subfigure}{0.328\textwidth}
\begin{tikzpicture}
    \node[] (pic) at (0,0) {\includegraphics[width=\textwidth]{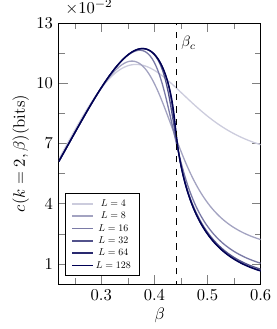}};
    \end{tikzpicture}
    \caption{Square lattice}
\end{subfigure}
\begin{subfigure}{0.328\textwidth}
\begin{tikzpicture}
    \node[] (pic) at (0,0) {\includegraphics[width=\textwidth]{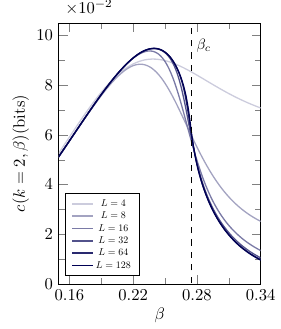}};
    \end{tikzpicture}
    \caption{Triangular lattice}
\end{subfigure}
\medskip
\begin{subfigure}{0.328\textwidth}
    \begin{tikzpicture}
    \node[] (pic) at (0,0) {\includegraphics[width=\textwidth]{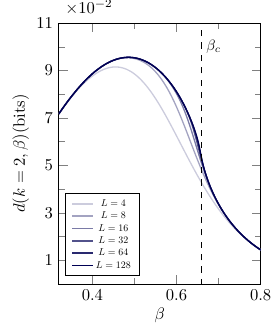}};
    \end{tikzpicture}
    \caption{Hexagonal lattice}
\end{subfigure}
\begin{subfigure}{0.328\textwidth}
\begin{tikzpicture}
    \node[] (pic) at (0,0) {\includegraphics[width=\textwidth]{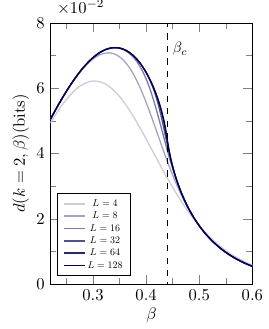}};
    \end{tikzpicture}
    \caption{Square lattice}
\end{subfigure}
\begin{subfigure}{0.328\textwidth}
\begin{tikzpicture}
    \node[] (pic) at (0,0) {\includegraphics[width=\textwidth]{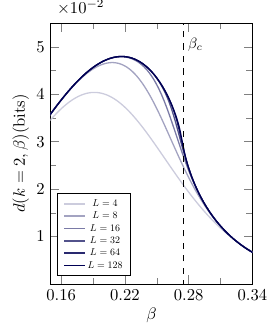}};
    \end{tikzpicture}
    \caption{Triangular lattice}
\end{subfigure}
\caption{The normalized pairwise complexity $c = C(k=2)/N$ and scale-specific shared information $d=D(k=2)/N$ as functions of $\beta$ for the (a),(d) hexagonal, (b),(e) square, and (c),(f) triangular lattices with $L=4, 8, 16, 32, 64$, and $128$. With the temperature resolution used ($\delta \beta = 0.001$), the error bars are represented as shaded regions around each curve, although not clearly visible.}
\label{Scale2Complexity}
\end{figure*}

We computed the normalized complexity $c(k=2)$ for the three lattices as a function of $\beta$ for $L = 4, 8, 16, 32, 64$ and $128$, as estimated from Monte Carlo simulations (see Sec. \ref{Subsection: ComputationalDetails}). The dashed lines in Fig.~\ref{Scale2Complexity} indicate the true critical temperatures $\beta_c$ of the lattices in the thermodynamic limit~\cite{baxter_exactly_2007}. Although not shown here, the complexity curves vanish as $\beta \rightarrow 0$ and asymptotically converge to unity in the strong-coupling regime. A global maximum is observed in the disordered phase for all system sizes. Remarkably, while the position of this maximum initially shifts closer to $\beta_c$ as the system size increases, it eventually stabilizes below $\beta_c$ for $L \geq 32$. This behavior contrasts with the critical behavior of the thermodynamic response functions, and the pairwise and global measures of mutual information, which have been shown to peak at the phase transition for the 2D Ising model~\cite{matsuda_mutual_1996,gu_universal_2007,barnett_information_2013}.
  
Nevertheless, several static and dynamic information-theoretic measures of classical spin models have been shown to exhibit a similar behavior to that of the complexity~\cite{wilms_mutual_2011,lau_information_2013,barnett_information_2013,SynergyWarning2019}. Specifically, the mutual information between two parts of a cylindrical 2D Ising lattice~\cite{wilms_mutual_2011,lau_information_2013}, and the global transfer entropy~\cite{barnett_information_2013} both exhibit peaks that reside in the disordered phase. The observed behavior has been linked to differing contributions of pairwise and higher-order correlation terms to these measures below $\beta_c$~\cite{barnett_information_2013}, with the divergence of their derivatives at $\beta_{c}^{'}$, as we will see at the end of this section, marking the onset of higher-order correlations $(k>2)$.

To examine the contribution of higher-order information correlations to the behavior of $C(k=2)$, using Eq.~\eqref{Equation: DFromC}, we computed the scale-specific information $d = D(k=2) / N$, normalized by $N$, which represents the shared information between two spins only, as shown in Figs.~\ref{Scale2Complexity}(d)–~\ref{Scale2Complexity}(f). Qualitatively, while the peak position of $d(k=2)$ is further away from $\beta_c$ than that of $c(k=2)$, its behavior is identical for all lattices. This suggests that the subcritical peaks are not strictly due to pairwise or higher-order correlations alone but result from a contribution of both components, namely $D(k=2)$ and $C(k=3)$. 

We investigate critical slowing down as a possible cause for the origin of the peaks of $C(k=2)$ and $D(k=2)$. It has been suggested in~\cite{barnett_information_2013} that a subcritical peak in a dynamic information-theoretic measure may serve as an early-warning signal, indicating that a system is approaching a critical transition from the disordered phase. This is, in part, due to the phenomenon of critical slowing down, where the relaxation time of the system to equilibrium increases as it approaches the transition point. In both stochastic and deterministic dynamics, this phenomenon manifests as increased autocorrelation times and variance in state variables~\cite{Scheffer2009}.

While critical fluctuations are an intrinsic property of the Ising model and are observed in any stochastic dynamics that samples from the Gibbs-Boltzmann distribution, critical slowing down is more specifically associated with local update dynamics such as the Metropolis algorithm. To verify that these subcritical peaks are not artifacts of this algorithm, we implemented computational techniques to reduce the effects of critical slowing down (see Appendix~\ref{Appendix: MCSimulation}), and repeated our simulations using Glauber dynamics~\cite{glauber1963time} with the same simulation parameters as in the Metropolis runs. Within the error bars, we found that the results were consistent, with their positions stabilizing in the disordered phase at the same temperatures for $L\geq 32$. 

In Figs.~\ref{Figure: LatticeComplexity}(a)–~\ref{Figure: LatticeComplexity}(c), the normalized complexity curves for adjacent, ascending system sizes ($L$, $2L$) exhibit approximate crossings near $\beta_c$, and for the temperature resolution used here ($\delta \beta = 0.001$), the crossing point is the same for the pairs ($32$, $64$) and ($64$, $128$). Additionally, the absence of similar crossings in $d(k=2)$ can be attributed to the exclusion of higher-order CMI terms that capture correlations beyond two spins. Such multispin correlations become increasingly relevant near criticality and are expected to exhibit scale-invariant behavior at $\beta_c$, which gives rise to the crossings of the normalized curves for different system sizes.

\begin{figure*}[ht!]
\centering
\begin{subfigure}{0.328\textwidth}
    \begin{tikzpicture}
    \node[] (pic) at (0,0) {\includegraphics[width=\textwidth]{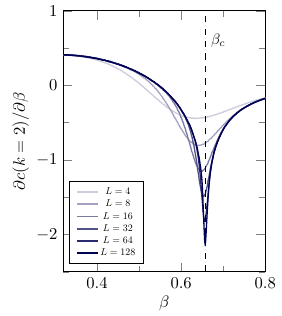}};
    \end{tikzpicture}
        \caption{Hexagonal lattice}
\end{subfigure}
\begin{subfigure}{0.328\textwidth}
\begin{tikzpicture}
    \node[] (pic) at (0,0) {\includegraphics[width=\textwidth]{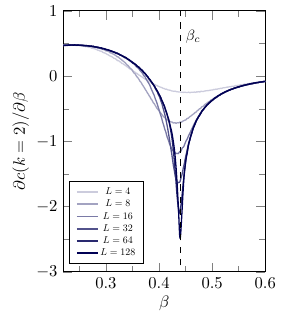}};
    \end{tikzpicture}
    \caption{Square lattice}
\end{subfigure}
\begin{subfigure}{0.328\textwidth}
\begin{tikzpicture}
    \node[] (pic) at (0,0) {\includegraphics[width=\textwidth]{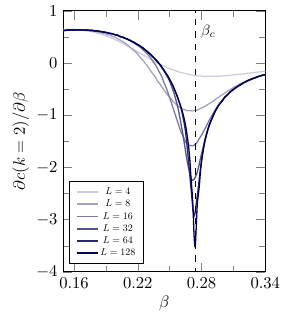}};
    \end{tikzpicture}
    \caption{Triangular lattice}
\end{subfigure}
\medskip
\begin{subfigure}{0.328\textwidth}
    \begin{tikzpicture}
    \node[] (pic) at (0,0) {\includegraphics[width=\textwidth]{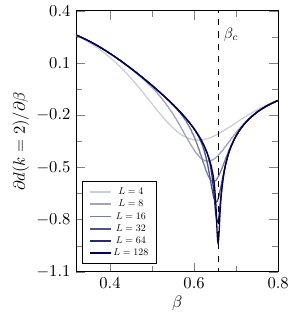}};
    \end{tikzpicture}
    \caption{Hexagonal lattice}
\end{subfigure}
\begin{subfigure}{0.328\textwidth}
\begin{tikzpicture}
    \node[] (pic) at (0,0) {\includegraphics[width=\textwidth]{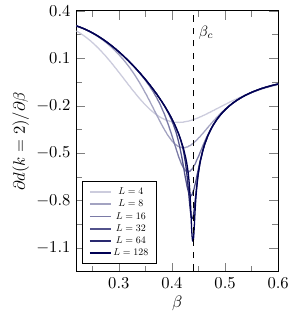}};
    \end{tikzpicture}
    \caption{Square lattice}
\end{subfigure}
\begin{subfigure}{0.328\textwidth}
\begin{tikzpicture}
    \node[] (pic) at (0,0) {\includegraphics[width=\textwidth]{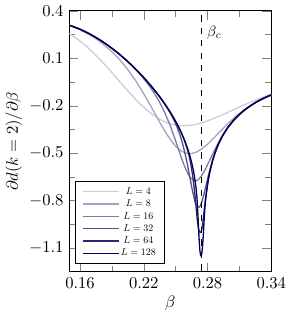}};
    \end{tikzpicture}
    \caption{Triangular lattice}
\end{subfigure}
\caption{The (numerical) derivatives of the normalized pairwise complexity and scale-specific shared information as functions of $\beta$ for the (a),(d) hexagonal, (b),(e) square and, (c),(f) triangular  lattices with $L=4, 8, 16, 32, 64$, and $128$. With the temperature resolution used ($\delta \beta = 0.001$), the error bars are represented as shaded regions around each curve, although not clearly visible.}
\label{Scale2ComplexityDerivatives}
\end{figure*}

We further investigate the critical behavior of the complexity curves in Fig.~\ref{Scale2Complexity} by analyzing their (numerical) temperature derivatives, $\partial c(k=2)/ \partial \beta$ and $\partial d(k=2)/ \partial \beta$, shown in Figs.~\ref{Scale2ComplexityDerivatives}(a)–~\ref{Scale2Complexity}(f). These derivatives appear to diverge in the vicinity of $\beta_c$, as indicated by the scaling of the amplitudes of their minima with the size of the system. A more detailed analysis of these derivatives shows that they diverge logarithmically at $\beta_c^{\prime}$ as
\begin{gather}\label{Equation: LogarithmicScaling}
\begin{alignedat}{1}
\min_{\beta} \left[\frac{1}{N}\frac{\partial C(k=2)}{\partial \beta}\right] &\sim A\ln{L^{-1}},\\[10pt]
\min_{\beta} \left[\frac{1}{N}\frac{\partial D(k=2)}{\partial \beta}\right] &\sim B\ln{L^{-1}},
\end{alignedat}
\end{gather}
where $A$ and $B$ are positive, lattice-dependent constants that differ between $C(k=2)$ and $D(k=2)$ (see Appendix~\ref{Appendix: CriticalExponents}). Moreover, these minima approach $\beta_c$ according to the power laws
\begin{gather}\label{Equation: PowerLawScaling}
\begin{alignedat}{1}
\beta_c - \arg \min_{\beta} \left[\frac{1}{N}\frac{\partial C(k=2)}{\partial \beta}\right] &\sim L^{-a},\\[10pt]
\beta_c - \arg \min_{\beta} \left[\frac{1}{N}\frac{\partial D(k=2)}{\partial \beta}\right] &\sim L^{-b}.
\end{alignedat}
\end{gather}

Both exponents $a$ and $b$ are found to be approximately equal to 1 for the three lattices (see Appendix~\ref{Appendix: CriticalExponents}), indicating that they may be related to the reciprocal of the correlation length critical exponent $\nu = 1$~\cite{baxter_exactly_2007}. Note that the uncertainties in both exponents are due to large corrections to scaling and the numerical estimation of the derivatives. Nevertheless, given that these are measures of information correlations, we speculate that these exponents are identical and universal, with their values across the three lattices confirming the universality class of the 2D Ising model. 

\section{Discussion}\label{Section: Discussion}
In this work, we evaluated the CP index as a tool for identifying the relevant DoF in 2D short-range, ferromagnetic Ising systems at different scales of observation. This approach allowed us to gain an information-theoretic perspective on the disorder-order phase transition. In particular, we report three main findings: (1) the CP captures phase-dependent multiscale structure that reflects key physical phenomena of the model, (2) the normalized pairwise measures $c(k=2)$ and $d(k=2)$ obey an area law away from criticality, and (3) the physical interpretation of the scale-specific shared information provides an insight into the mathematical definition of the CP in terms of the exclusive shared information at scale $k$ or higher.

We found that the CP characterizes the Ising model as a simple system in both the high-temperature (paramagnetic) and low-temperature (ferromagnetic) regimes, where scale-dependent structures are largely absent. This is in contrast with traditional measures of complexity such as thermodynamic or information entropies, which reach their extremal values in these regimes, and may, therefore, suggest that a system is either more or less complex. In the critical region, however, multiscale structure emerges, with the systems exhibiting distinct behaviors at different scales. Notably, the emergence of macroscopic magnetization in the ferromagnetic phase is marked by oscillations in the extremal values of the scale-specific shared information $D_{\text{ex}}(k)$ at large scales. 

A key signature of magnetic domain formation is the appearance of extremal values in $D(k)$ at characteristic temperatures for each scale $k$. We interpret these temperatures as those at which magnetic domains of size $k$ spins are predominantly formed. The reason for this is twofold. First, each extremum indicates that information correlations at scale $k$ are strongest at that temperature, where the \textit{exclusive} shared information within subsystems of size $k$ spins is maximized. Second, due to the Markovian nature of finite-range Ising models, the CMI satisfies the equality
\begin{equation}\label{ZeroConditionalMutualInformation}
I(\sigma_1^{(u)}; \ldots; \sigma_{k}^{(u)} | \partial_1^{(u^{\prime})}, \ldots, \partial_{N-k}^{(u^{\prime})}) = 0,
\end{equation}
if at least one spin in $\boldsymbol{\sigma}^{(u)} = \{\sigma_1^{(u)}, \ldots ,\sigma_{k}^{(u)}\}$ is conditionally independent of the others, given the configuration of the separating boundary $\boldsymbol{\partial}^{(u^{\prime})} =\{\partial_1^{(u^{\prime})}, \ldots, \partial_{N-k}^{(u^{\prime})}\}$. This condition implies that only subsystems that form connected spin clusters have non-vanishing contributions to $C(k)$ and $D(k)$. Since $D_{\text{ex}}(k)$ is at finite temperatures for $k\geq 2$, it corresponds to the case when a separating boundary provides little to no information about a spin cluster, while the $k$ spins within it are strongly correlated. This is characteristic of magnetic domains separated by boundaries that exhibit local disorder relative to the clusters.  

For larger system sizes, the measures $C(k=2)$ and $D(k=2)$ share universal behaviors, as evidenced by their normalized forms and corresponding temperature derivatives shown in Figs.~\ref{Scale2Complexity} and \ref{Scale2ComplexityDerivatives}. Specifically, these quantities are extensive away from $\beta_c$, while their derivatives diverge logarithmically with the size of the system at criticality. Moreover, the height and width of their maxima are independent of the system size for $L \geq 32$, strongly suggesting that they remain finite in the thermodynamic limit ($L\rightarrow \infty$), despite the divergence of the correlation length at $\beta_c$. 

Given that both measures are functions of the CMI, which can be expressed as a combination of Shannon entropies as in Eq.~\eqref{Equation: ConditionalMutualInformation}, their extensivity implies that an area law is obeyed. This is because in classical and quantum lattice models, the area law states that at finite temperatures, the mutual information between two subsystems is bounded by the entropy of the interface separating them~\cite{wolf_area_2008}. 

For a Markov random field, this result is not surprising, as the number of DoF in the conditional entropy reduces to that of the boundary, leading to localized information correlations between subsystems at this boundary. However, $C(k=2)$ includes a co-information term~\cite{Timme2014}, $D(k=N)$, which is a global measure corresponding to the unconditional mutual information among the $N$ spins. In contrast to other global measures~\cite{barnett_information_2013}, this term does not lead to the divergence of $C(k=2)$ at $\beta_c^{\prime}$.

Our final main finding is concerned with the theoretical basis for the definition of the CP, as given in Eq.~\eqref{Equation: ComplexityProfile}, which is rooted in the condition expressed in Eq.~\eqref{ZeroConditionalMutualInformation}. The hierarchical nature of the CP reflects the progressive exclusion of contributions from smaller-scale CMI terms as the scale increases. While the exclusive shared information within a strongly correlated subsystem can be zero, this does not imply that the spins are uncorrelated. Rather, it indicates that they are part of a larger correlation mediated by the separating boundary. To capture these correlations and provide an accurate representation of shared information, the contributions from scale $k$ and higher are incorporated into the CP, which account for the effective DoF at scale $k$. 

This remark is particularly relevant when considering how higher-order correlations impact the effective number of DoF in a Markov random field. For a system of $N$ noninteracting spins, its microstate can be fully described by $N$ independent spin variables, each contributing one bit of entropy. In the presence of correlations, however, spin subsystems may behave collectively, giving rise to structure that reduces the number of independent variables needed to describe the state of a system at different scales. This is demonstrated by the conservation of the sum rule of the CP (see Sec.~\ref{Subsection: ComplexityVsScale}). The CP therefore identifies these relevant DoF by quantifying how statistical dependencies, measured here via the CMI, are distributed across scales. A non-zero value of $C(k)$ indicates that subsystems of $k$ spins exhibit significant shared information at scale $k$ or higher, effectively acting as collective units or higher-scale DoF. Such correlated groups can thus be treated as emergent entities, which, in the spatial Ising model, correspond to magnetic domains, whose scale is the sum of the scales of their parts. In this way, the CP reveals the scales at which collective behaviors emerge and carry irreducible or characteristic dependencies.

We note that while the multiscale complexity formalism is a powerful framework for providing a quantitative measure for scale-dependent complexity and identifying the relevant DoF in complex systems, its practical application is limited by significant computational challenges. As defined, evaluating the CP requires computing multivariate information-theoretic quantities, which becomes computationally prohibitive for large systems for any choice of the information function. In Markov random fields, where the system exhibits local interactions and satisfies the Markov property, this complexity can be greatly reduced by leveraging known symmetries and locality constraints. However, in more general or unfamiliar systems, such properties may be unknown or unjustified. Consequently, the applicability of the CP to broader classes of complex systems remains limited without additional constraints or approximations.

Despite these practical challenges, the present study demonstrates the applicability of the CP to equilibrium systems with finite-range interactions and classical Gibbs distributions, where the conditional information measures of Markov random fields are exploited. These systems include, for example, the $q$-state Potts model and continuous-spin systems such as the \textit{XY} and Heisenberg models. 
Deriving explicit subsystem probability distributions analogous to Eq.~\eqref{Equation: MarginalProbabilityDistribution} is, however, mathematically demanding for continuous-variable systems. More broadly, nonequilibrium systems may be treated probabilistically by constructing empirical subsystem distributions from long-time histories. Such estimates may over- or underestimate the information content of specific subsets since the state space is partially accessible in finite-time simulations.
\section{Conclusions}\label{Section: Conclusion}
In conclusion, by using the CP to analyze the multiscale behavior of spatial Ising systems with ferromagnetic interactions, we have demonstrated its ability to identify the scales at which collective behaviors emerge, revealing the relevant DoF that govern the behavior of the system at different levels of organization. We established a correspondence between the information-theoretic behaviors captured by the CP and key physical phenomena across temperature regimes, including disorder, magnetic domain formation, and the emergence of macroscopic magnetization. We remark that the subcritical peak in the pairwise complexity may reflect a universal phenomenon of certain classes of second-order phase transitions, as captured by measures of general dependence applied to the 2D Ising model~\cite{wilms_mutual_2011,lau_information_2013,barnett_information_2013,SynergyWarning2019}. In future work, we aim to investigate this potential universality and understand its physical origins by applying the measure to other systems within the same universality class, as well as to higher-dimensional systems.

The simulation scripts used to compute $C(k=2)$ and $D(k=2)$ for large systems, as well as the binary expressions of the multipin-coding algorithm (see Appendix~\ref{Appendix: MSC}) for each lattice, are available on Zenodo~\cite{al_azki_2025_15515297}.
\begin{acknowledgments}
This research was supported by an Australian Government Research Training Program (RTP) Scholarship. Computational resources and services were provided by the National Computational Infrastructure (NCI), which is supported by the Australian Government. The authors thank Professor Jared Cole for insightful discussions and Dr. Francesco Campaioli for carefully reading the manuscript. 
\end{acknowledgments}

\appendix
\section{Mathematical derivations}
\subsection{Marginal probability distribution}\label{Appendix: MarginalProabilityDistribution}
The probability of observing a spin subsystem in configuration $\sigma^{(u)}=(\sigma_{1}^{(u)}, \ldots, \sigma_k^{(u)})$ is obtained by summing over the probabilities of the configurations of the system that correspond to this subsystem configuration, as in Eq.~\eqref{Equation: ExampleMarginalProbability}. Formally, for a subsystem $\boldsymbol{\sigma}^{(u)}=\{\sigma_{1}^{(u)}, \ldots, \sigma_{k}^{(u)}\}$ with $k$ spins, we can express this marginal probability as
\begin{equation}\label{Equation: MarignalConditional}
    p(\sigma_{1}^{(u)}, \ldots, \sigma_k^{(u)}) = \sum_{s} p(s)p(\sigma^{(u)}=s^{(u)}|s),
\end{equation}
where given the configuration $s$ of the system, the conditional probability $p(\sigma^{(u)}=s^{(u)}|s)$ is $1$ if all states of the spins in the subsystem correspond to those in the system and $0$ otherwise. Using the identity
\begin{equation}\label{Equation: SpinIdentity}
    \delta(\sigma_i^{(u)}, s_i^{(u)}) = \frac{1}{2}(1+\sigma_i^{(u)}s_i^{(u)}),
\end{equation}
the conditional probability is then written as
\begin{align}
    p(\sigma^{(u)}=s^{(u)}|s) &=  \delta(\sigma^{(u)}_1, s^{(u)}_1)\delta(\sigma^{(u)}_2, s^{(u)}_2)\cdots \nonumber\\[5pt] 
    &\quad \times \delta(\sigma^{(u)}_k, s^{(u)}_k).
\end{align}
Substituting in Eq.~\eqref{Equation: MarignalConditional}, we have
\begin{align}\label{Equation: MarginalProbabilityExpanded}
p(\sigma_{1}^{(u)}, \ldots, \sigma_k^{(u)}) &= \sum_{s} p(s) \delta(\sigma^{(u)}_1, s^{(u)}_1)\cdots \delta(\sigma^{(u)}_k, s^{(u)}_k)\nonumber\\[5pt]
&= \frac{1}{2^k}\sum_{s}p(s)(1+\sigma^{(u)}_1 s^{(u)}_1)\cdots \nonumber \\[5pt]
&\quad \times (1+\sigma^{(u)}_k s^{(u)}_k).
\end{align}
By expanding the right-hand side, we will now refer to the combination of $l$ spins from subsystem $\boldsymbol{\sigma}^{(u)}$ by the superscript $v$ as follows:
\begin{align}
        p(\sigma_{1}^{(u)}, \ldots, \sigma_k^{(u)}) &= \frac{1}{2^k}\sum_{s}p(s)\Big(1+\sigma^{(1)}_1 s^{(1)}_1+\cdots \nonumber \\[5pt]
                                                    &\quad +\sigma^{(v)}_1\cdots\sigma^{(v)}_k s^{(v)}_1\cdots s^{(v)}_k\Big) \nonumber\\[5pt]
                                                    &= \frac{1}{2^k}\Big(1+\sigma^{(1)}_1\langle s^{(1)}_1\rangle_{\mathcal{H}}+\cdots \nonumber \\[5pt]
                                                    &\quad +\sigma^{(v)}_1\cdots\sigma^{(v)}_k \langle s^{(v)}_1\cdots s^{(v)}_k\rangle_{\mathcal{H}}\Big).
\end{align}
Given that the odd-point correlation functions vanish in the absence of an external field, we have the general expression for the subsystem probability distribution
\begin{align}
p(\sigma_{1}^{(u)}, \ldots, \sigma_k^{(u)}) &= \frac{1}{2^k}\bigg(1+\sum_{\ell\text{ even}}^k \sum_{\boldsymbol{\sigma}^{(v)}} \sigma^{(v)}_1\cdots\sigma^{(v)}_{\ell} \nonumber \\[5pt]
&\quad \times \langle s_1^{(v)}\cdots s_{\ell}^{(v)}\rangle_{\mathcal{H}}\bigg),
\end{align}
where the inner summation is over all possible subsets $\boldsymbol{\sigma}^{(v)}\subseteq\{\sigma_1^{(u)}, \ldots, \sigma_{k}^{(u)}\}$ of size $\ell$.

\subsection{Conditional probability}\label{Appendix: ConditionalProbability}
Let $\boldsymbol{\sigma}^{(u)}$ denote a subsystem comprising $k$ spins, and $\boldsymbol{\sigma}^{(u')}$ its complement with $N-k$ spins, in a system $\boldsymbol{s}$ of total size $N$. The Hamiltonian in Eq.~\eqref{Equation: IsingHamiltonian} can then be partitioned as
\begin{equation}\label{HamiltonianPartition}
    \mathcal{H}(s) = \mathcal{H}_A + \mathcal{H}_B,
\end{equation}
where $\mathcal{H}_A$ includes all interaction terms between spins within subsystem $\boldsymbol{\sigma}^{(u)}$ as well as interactions between spins in $\boldsymbol{\sigma}^{(u)}$ and its complement $\boldsymbol{\sigma}^{(u')}$. In contrast, $\mathcal{H}_B$ consists only of interaction terms between spins entirely within subsystem $\boldsymbol{\sigma}^{(u')}$. From the definition of the conditional probability, the probability of observing configuration $\sigma^{(u)}$ conditioned on the configuration of subsystem $\boldsymbol{\sigma}^{(u^{\prime})}$ is given by
\begin{gather}
    \begin{alignedat}{1}
        p(\sigma^{(u)} | \sigma^{(u^{\prime})}) &= \frac{p(\sigma^{(u)}, \sigma^{(u^{\prime})})}{p(\sigma^{(u^{\prime})})}\\[10pt]
                     &= \frac{p(s)}{p(\sigma^{(u^{\prime})})},
    \end{alignedat}
\end{gather}
where $p(\sigma^{(u)}, \sigma^{(u^{\prime})}) = p(s)$ is the joint system distribution, as defined in Eq.~\eqref{Equation: BoltzmannDistribution}, and the marginal distribution $p(\sigma^{(u^{\prime})})$ is given by Eq.~\eqref{Equation: MarginalProbabilityExpanded}. Substituting these expressions, we have
\begin{gather}\label{Equation: ConditionalProbability}
    \begin{alignedat}{1}
        p(\sigma^{(u)} | \sigma^{(u^{\prime})}) &= \frac{Z^{-1} e^{-\beta \mathcal{H}(s)}}{\sum_s p(s) \delta(\sigma^{(u^{\prime})}_1, s^{(u^{\prime})}_1)\cdots \delta(\sigma^{(u^{\prime})}_{N-k}, s^{(u^{\prime})}_{N-k})} \\[10pt]
                     &= \frac{Z^{-1} e^{-\beta \mathcal{H}(s)}}{\sum_{s^{(u)}} p(s)} \\[10pt]
                     &= \frac{Z^{-1} e^{-\beta \mathcal{H}(s)}}{\sum_{s^{(u)}}Z^{-1} e^{-\beta \mathcal{H}(s)}} \\[10pt]
                     &= \frac{Z^{-1} e^{-\beta (\mathcal{H}_A + \mathcal{H}_B)}}{\sum_{s^{(u)}} Z^{-1} e^{-\beta (\mathcal{H}_A + \mathcal{H}_B)}}\\[10pt]
                     &= \frac{Z^{-1} e^{-\beta \mathcal{H}_B} e^{-\beta \mathcal{H}_A}}{Z^{-1} e^{-\beta \mathcal{H}_B} \sum_{s^{(u)}} e^{-\beta \mathcal{H}_A}}\\[10pt]
                     &= \frac{e^{-\beta \mathcal{H}_A}}{\sum_{s^{(u)}} e^{-\beta \mathcal{H}_A}},
    \end{alignedat}
\end{gather}
where the summation in the denominator is over the configurations of subsystem $\boldsymbol{\sigma}^{(u)}$. We note that this expression generalizes the one-variable conditional probability commonly used in Gibbs sampling~\cite{MurphyKevinP.2012Ml:a}, where each variable is sampled conditioned on its boundary in an undirected graphical model, to subsystems of arbitrary sizes.

\section{Monte Carlo simulations}\label{Appendix: MCSimulation}
For large systems, we implemented two Monte Carlo schemes to compute the pairwise complexity $C(k=2)$ and the scale-specific shared information $D(k=2)$ in Fig.~\ref{Scale2Complexity}. Each scheme is used to estimate one of the two terms of the subsystem entropy in Eq.~\eqref{Equation: JointEntropy}: the joint entropy and the reduced conditional entropy. The joint system entropy is computed using the replica-exchange Wang-Landau (REWL) algorithm, based on the estimated density of states (DoS), while the conditional entropy is computed using the conditional probability $p(\sigma^{(u)}|\partial ^{(u^{\prime})})$ in Eq.~\eqref{Equation: ConditionalProbability} and the marginal probability distribution $p(\partial^{(u^{\prime})})$, whose correlation functions are estimated using a multispin-coded Metropolis algorithm.

\subsection{Replica-exchange Wang-Landau algorithm}\label{Appendix: REWL}
We estimated the DoS, $g(E)$, for the three lattices using the REWL algorithm~\cite{vogel_scalable_2014,PhysRevLett.Vogel-Generic}. In a standard Wang-Landau sampling~\cite{WLSampling}, a single random walker performs a biased walk in the energy space, updating the DoS [$\ln g(E) \rightarrow \ln g(E) + \ln f_i$] by a modification factor $f_0=e^1$ that is iteratively reduced, and incrementing the histogram $n(E)$ at each visited energy state. The acceptance probability for transitioning from configuration $\mu$ to configuration $\nu$ is given by
\begin{equation}\label{Equation: Wang-LandauCriteria}
    p(\mu \rightarrow \nu)=\min{\left\{1, \exp{\left[\ln g(E_{\mu}) - \ln g(E_{\nu})\right]}\right\}}.
\end{equation}
The histogram flatness criterion is evaluated every $10^4$ lattice sweeps to check whether all energy states have been sampled approximately equally. Specifically, the histogram $n(E)$ is considered \textit{flat} if all entries satisfy $n(E) \geq 0.9 \langle n(E) \rangle$, where $\langle n(E) \rangle$ denotes the average count of the energy histogram. Once the criterion is met, the algorithm proceeds to the next iteration by resetting the histogram to $n(E) = 0$ for all $E$ and reducing the modification factor according to the standard function $\ln f_{i+1} = \ln \sqrt{f_i}$. The simulation terminates when $\ln f_i < 10^{-8}$, and the final DoS is normalized using the known ground-state degeneracy, $g(E_{\text{min}}) = 2$.

To improve scalability for large systems and allow faster convergence in systems with an asymmetric DoS, we employ the REWL algorithm. The energy space is partitioned into overlapping windows with $50\%$ overlap, each assigned local walkers that perform independent Wang–Landau updates. Replica exchanges between adjacent windows $j$ and $j+1$ are attempted every $10^3$ lattice sweeps and are accepted with probability
\begin{equation}\label{Equation: ReplicaExchangeWangLandauCriterion}
    p(\mu \leftrightarrow \nu)= \min \left\{1, \frac{\exp\left[\ln g_j(E_{\mu}) + \ln g_{j+1}(E_{\nu})\right]}{\exp\left[\ln g_j(E_{\nu}) + \ln g_{j+1}(E_{\mu})\right]} \right\}.
\end{equation}
After convergence, the local DoS estimates are merged using a derivative-matching procedure to ensure continuity~\cite{vogel_scalable_2014}. For implementation details, including message passing interface-based parallelization for distributed computation, we refer the reader to~\cite{Vogel_2018}. Thermodynamic observables, including the Gibbs entropy, are then computed by estimating the partition function via
\begin{equation}
        \ln \mathcal{Z} = \ln{\left[\sum_{E} e^{\ln{g(E)}-\beta E - \lambda}\right]} + \lambda,
\end{equation}
where $\lambda=\arg \max_{E} \left[\ln{g(E) - \beta E}\right]$ is the largest exponent, introduced to prevent numerical overflow when evaluating exponentials with large arguments.

\subsection{Multispin coding}\label{Appendix: MSC}
Multispin coding is a technique that enables the parallel simulation of independent binary-spin systems by encoding their states into the bits of a single machine word~\cite{PhysRevLett.42.1390}. On a $64$-bit architecture, we implement an asynchronous update~\cite{newman1999monte} of the Metropolis algorithm to obtain a large number of samples for the spin correlation functions. In this approach, the $64$ bits of a word simultaneously represent the states of a single spin $s_i$ in $64$ independent replicas of the system at the same temperature. 

\subsubsection{Metropolis algorithm}
We use the Metropolis single spin-flip algorithm to simulate the Ising model~\cite{newman1999monte}, in which each spin $s_i$ is randomly selected and flipped with probability
\begin{equation}\label{Equation: MetropolisCriteria}
    p(\mu \rightarrow \nu)= \min\left[1, \exp(-\beta \Delta\mathcal{H})\right].
\end{equation}
The change in energy between the two configurations is given by
\begin{equation}\label{Equation: TransitionEnergy}
    \Delta\mathcal{H}= 2Js_i \sum_j s_j,
\end{equation}
with the sum being over the nearest neighbors of the selected spin $i$. For the lattice geometries considered in this paper (see Fig.~\ref{Figure: LatticeDefinitions}), the possible values of $\Delta\mathcal{H}$ are as follows:
\begin{align}\label{Equation: EnergyChanges}
    \Delta\mathcal{H}_{\text{Hex}} &\in \{-6J, -2J, 2J, 6J\}, \nonumber\\
    \Delta\mathcal{H}_{\text{Sqr}} &\in \{-8J, -4J, 0J, 4J, 8J\}, \\
    \Delta\mathcal{H}_{\text{Tri}} &\in \{-12J, -8J, -4J, 0J, 4J, 8J, 12J\}. \nonumber
\end{align}

\subsubsection{Asynchronous multispin-coded Metropolis algorithm.}
We represent the spin states of site $i$ across 64 replicas as a single $64$-bit integer $\texttt{S}_i$, whose binary representation is $\text{bin}(\texttt{S}_i)=\texttt{s}_1\texttt{s}_2\cdots\texttt{s}_{64}$. Each bit $\texttt{s}_r$ corresponds to the spin value $s_r$ in the $r$-th replica, with the following mapping:
\begin{align*}
   s_r = -1 &\;\rightarrow\; \texttt{s}_r = 0, \\
   s_r = +1 &\;\rightarrow\; \texttt{s}_r = 1.
\end{align*}
In the Metropolis algorithm, a spin is always flipped when at least $\lceil z/2 \rceil$ of its neighbors are anti-aligned ($\Delta\mathcal{H}\leq 0$). Otherwise, for exactly $n$ anti-aligned neighbors, the spin is flipped with the respective transition probability $p_n(\mu \rightarrow \nu)$ determined by the energy change in Eq.~\eqref{Equation: EnergyChanges}. The multispin-coded Metropolis update is implemented via the \textit{master expression}
\begin{equation}
 \texttt{S}_i^{'} = \texttt{S}_i \oplus \left[R_{\geq {\lceil z/2\rceil}} \vee \cdots\vee \left(R_{\geq 1} \wedge r_{1}\right) \vee r_{0} \right],
\end{equation}
where $\wedge, \vee, \oplus$ denote the bitwise AND, OR, and XOR operations, respectively. The terms are defined as follows:
\begin{itemize}
    \item[(i)] $\texttt{S}_i^{'}$ is the updated $64$-bit integer at site $i$.
    \item[(ii)] $r_n$ is a random $64$-bit integer in which each bit is independently set to \texttt{1} with probability $p_n$.
    \item[(iii)] $R_{\geq n}$ is a $64$-bit mask in which each bit is set to \texttt{1} if the spin at site $i$ in the corresponding replica has $n$ or more anti-aligned neighbors.
\end{itemize}
It follows logically that if the $r$th bit in $R_{\geq n}$ is set to \texttt{1}, then the $r$th bits of all masks $R_{\geq n-1}, \ldots, R_{\geq 1}$ must also be set to \texttt{1}. Therefore, the probability that the composite logical expression
\begin{align*}
    \left[\left(R_{\geq n} \wedge r_{n}\right) \vee \cdots \vee r_{0}\right]
\end{align*}
evaluates to \texttt{1} corresponds to the transition probability $p_n(\mu \rightarrow \nu)$ when exactly $n$ neighbors are anti-aligned. Defining $p_n$ as the probability that the bits of $r_n$ are set to \texttt{1}, the probability that none of the bits in $r_0, \ldots, r_n$ are \texttt{1} can be expressed as
\begin{equation}
    1 - p_n(\mu \rightarrow \nu) = \prod_{i=0}^n (1 - p_i).
\end{equation}
From this, the individual probabilities $p_n$ are derived recursively, starting with $p_0$ and using each previously computed value to determine the next, until all are obtained. 

To derive the expressions for $R_{\geq n}$, we first define the following intermediate variables:
\begin{equation}
    a_j = \texttt{S}_i \oplus \texttt{S}_j^{(i)} \quad \text{for } j = 1, \ldots, z,
\end{equation}
where $a_j$ is a $64$-bit integer whose bits are set to \texttt{1} if the nearest-neighbor pair $\left<i,j\right>$ is anti-aligned in the corresponding replica. The bit-mask $R_{\geq n}$ is then defined as
\begin{equation}
    R_{\geq n} =
\bigvee_{\substack{T \subseteq \{a_1, \ldots, a_z\} \\ |T| = n}} 
\left( \bigwedge_{a_j \in T} a_j \right)
\end{equation}
where $T$ is any subset of the $z$ variables $a_j$ of size $n$. This sets a bit to \texttt{1} if at least $n$ of its neighbors are anti-aligned. The full expressions for $\texttt{S}_i^{'}$, $R_{\geq n}$ and $p_n$ for each of the three lattice geometries are provided in the supplemental material available on Zenodo~\cite{al_azki_2025_15515297}.

\subsubsection{Pseudorandom bit generation}
We implement the Poisson-Or algorithm~\cite{RandomBitWatanabe} to generate the random-bit variables $r_n$, without setting the bits individually. The algorithm models $r_n$ as the bitwise OR of a random number $k$ of integers, each with exactly one bit set to \texttt{1} (uniformly chosen among $64$ bits). A Poisson process is used to model the generation of these $k$ integers, where $k \sim \text{Poisson}(\lambda)$. The probability that the bits in $r_n$ are set to \texttt{1} is then given by
\begin{equation}
 p_n = 1 - \exp(-\lambda / 64).   
\end{equation}
Setting $\lambda = -64 \ln(1 - p_n)$ ensures that the bits are efficiently set to $\texttt{1}$ with the desired probability $p_n$. A detailed implementation of the algorithm is provided in~\cite{RandomBitWatanabe}.

\subsubsection{Simulation setup}
We simulated the Ising model using the previously described multispin-coded Metropolis algorithm to estimate the correlation functions associated with the probability distribution $p(\partial^{(u^{\prime})})$, for system sizes $L=4, 8, 16, 32, 64$ and $128$. To accurately locate the maxima of $C(k=2)$ and $D(k=2)$, and the minima of their derivatives, we used a temperature resolution of $\delta\beta=0.001$. The simulations started in the high-temperature regime with random initializations of the spin configurations. For each temperature, the system was equilibrated with $10^4$ lattice sweeps, after which measurements were performed. The final configuration at each temperature was used as the initial configuration for the next (lower) temperature step.

Taking advantage of the multispin-coded algorithm, we performed $40$ independent simulations for each correlation function estimation at each temperature, each with $20\times10^6 \times 64$ total number of measurements. Due to critical slowing down near the phase transition, subsequent measurements are highly correlated. To ensure statistical independence, the autocorrelation time $\tau$ was estimated before sampling the correlation functions~\cite{newman1999monte}. Measurements were then taken every $2\tau$ lattice sweeps. As a result, the total simulation time varied with both temperature and system size. The simulation script is available on Zenodo~\cite{al_azki_2025_15515297}.

\begin{widetext}
\section{Supplementary data}
\subsection{Full-scale complexity $(N=20)$}\label{Appendix: FullScaleComplexity}
In Fig.~\ref{Figure: LatticeComplexity20}, we present the exact CPs for the square and triangular lattices with $N=20$. For the square lattice, we notice the emergence of negative complexity at scale $k=19$, which was not observed for $N=18$ [see Fig.~\ref{Figure: LatticeComplexity}(b)]. For the triangular lattice, the amplitude of the negative minimum at scale $k=19$ is larger than that observed at scale $k=17$ for $N=18$ [see Fig.~\ref{Figure: LatticeComplexity}(c)]. Figure~\ref{Figure: LatticeComplexity20}(c) shows the scale-specific shared information at scale $k=19$ with a negative minimum at $\beta=0.368$. Figure~\ref{Figure: LatticeComplexity20}(d) illustrates the competing effects of thermal fluctuations and long-range order at the same scale. Thermal fluctuations are quantified by the variance of the correlation function of a 19-spin subset, whereas long-range order is characterized by corresponding correlation function itself. We demonstrate the competing effects for $D(k=19)$ instead of $C(k=19)$ since the latter includes correlations at this scale and higher, while the correlation function captures correlations among exactly $19$ spins, consistent with the definition of the former [see Eq.~\eqref{Equation: ScaleSpecificInformation}]. Due to the permutational symmetry at this scale, it is sufficient to the consider the correlation function of a single subset. In addition, a restricted sum of the partition function is performed to obtain a nonzero odd-point correlation function. At $\beta=0.368$, we can see that the two effects are balanced, which causes a system simulated using single spin-flip dynamics to transition between magnetized states with 19 aligned spins along the positive and negative macroscopic orientation.

\begin{figure*}[!ht]
\centering
\begin{subfigure}{0.496\textwidth}
    \begin{tikzpicture}
    \node[] (pic) at (0,0) {\includegraphics[width=\textwidth]{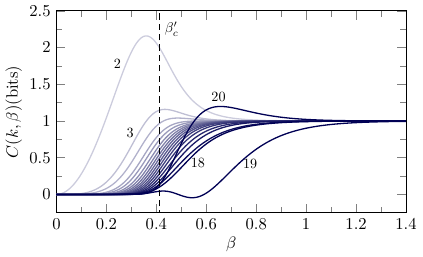}};
    \end{tikzpicture}
    \caption{Square lattice}
\end{subfigure}
\hfill
\begin{subfigure}{0.496\textwidth}
\begin{tikzpicture}
    \node[] (pic) at (0,0) {\includegraphics[width=\textwidth]{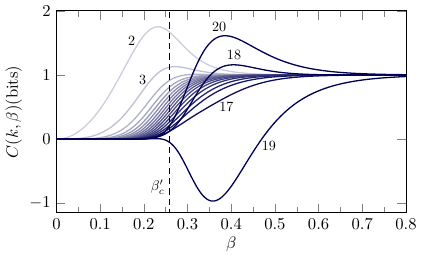}};
    \end{tikzpicture}
    \caption{Triangular lattice}
\end{subfigure}
\medskip
\begin{subfigure}{0.496\textwidth}
    \begin{tikzpicture}
    \node[] (pic) at (0,0) {\includegraphics[width=\textwidth]{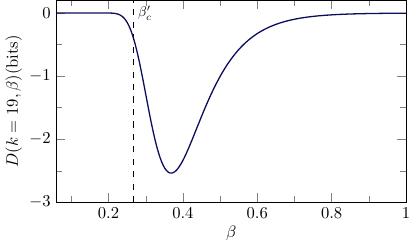}};
    \end{tikzpicture}
    \caption{Triangular lattice ($k=19$)}
\end{subfigure}
\hfill
\begin{subfigure}{0.496\textwidth}
\begin{tikzpicture}
    \node[] (pic) at (0,0) {\includegraphics[width=\textwidth]{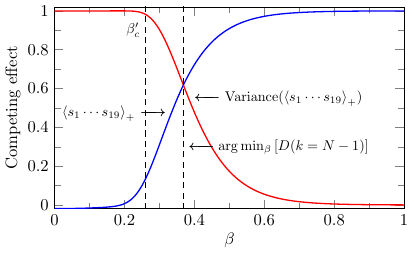}};
    \end{tikzpicture}
    \caption{Triangular lattice ($k=19$)}
\end{subfigure}
\caption{The complexities as functions of $\beta$ for the ferromagnetic (a) square and (b) triangular lattices with $N=20$ spins. The curves are labeled by scale from $k=2$ (light) to $k=20$ (dark). (c) The negative scale-specific shared information for the triangular lattice at scale $k=19$, and (d) the corresponding competing effects of thermal fluctuations and long-range order. These effects are represented, respectively, by the variance of the correlation function and the correlation function itself for a single subset at scale $k=19$. Since scale $k=19$ has an odd parity under a spin flip, the correlation function is evaluated in the positive-magnetization Gibbs state, denoted by $\left<\cdot\right>_+$, which is equivalent to taking the zero-field limit $h\rightarrow0^+$.}
\label{Figure: LatticeComplexity20}
\end{figure*}

\subsection{Scaling analysis}\label{Appendix: CriticalExponents}
The scaling analysis of the derivatives of the measures $c(k=2)$ and $d(k=2)$ is shown in Fig.~\ref{Figure: ScalingAnalysis}. The logarithmic and power-law scaling relations follow the forms given in Eqs.~\eqref{Equation: LogarithmicScaling} and~\eqref{Equation: PowerLawScaling}. The estimated constants $A$ and $B$, the exponents $a$ and $b$, and the estimated reciprocal of the correlation length exponent associated with specific heat for each lattice are summarized in Table~\ref{Table: Constants and Exponents}.

\begin{figure*}[!ht]
\centering
\begin{subfigure}{0.474\textwidth}
    \begin{tikzpicture}
    \node[] (pic) at (0,0) {\includegraphics[width=\textwidth]{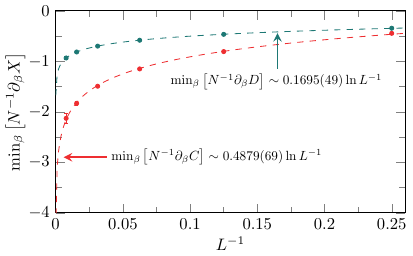}};
    \end{tikzpicture}
    \caption{Hexagonal lattice}
\end{subfigure}
\hfill
\begin{subfigure}{0.496\textwidth}
\begin{tikzpicture}
    \node[] (pic) at (0,0) {\includegraphics[width=\textwidth]{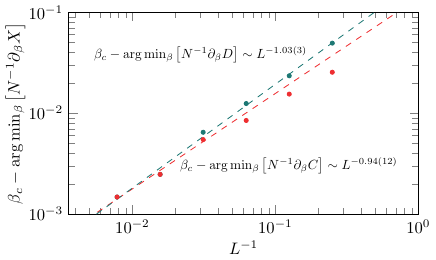}};
    \end{tikzpicture}
    \caption{Hexagonal lattice}
\end{subfigure}
\medskip
\begin{subfigure}{0.474\textwidth}
\begin{tikzpicture}
    \node[] (pic) at (0,0) {\includegraphics[width=\textwidth]{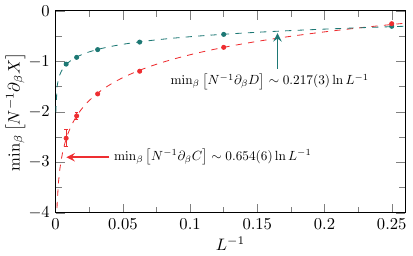}};
    \end{tikzpicture}
    \caption{Square lattice}
\end{subfigure}
\hfill
\begin{subfigure}{0.496\textwidth}
\begin{tikzpicture}
    \node[] (pic) at (0,0) {\includegraphics[width=\textwidth]{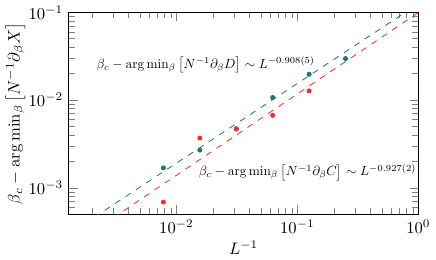}};
    \end{tikzpicture}
    \caption{Square lattice}
\end{subfigure}
\medskip
\begin{subfigure}{0.474\textwidth}
\begin{tikzpicture}
    \node[] (pic) at (0,0) {\includegraphics[width=\textwidth]{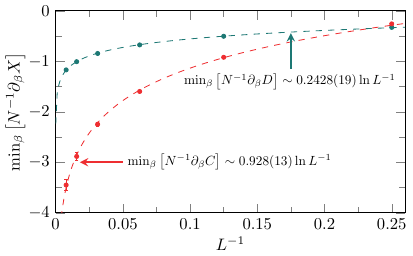}};
    \end{tikzpicture}
    \caption{Triangular lattice}
\end{subfigure}
\hfill
\begin{subfigure}{0.496\textwidth}
\begin{tikzpicture}
    \node[] (pic) at (0,0) {\includegraphics[width=\textwidth]{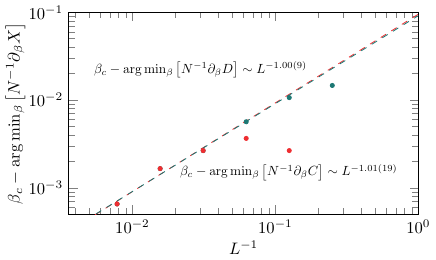}};
    \end{tikzpicture}
    \caption{Triangular}
\end{subfigure}
\caption{(a),(c),(e) Logarithmic scaling of the amplitudes of the minima of the derivatives of \( c(k=2) \) and \( d(k=2) \), and (b),(d),(f) log-log plots of the positions of these minima relative to the true critical temperatures \( \beta_c \), as functions of the inverse system size for the (a),(b) hexagonal, (c),(d) square, and (e),(f) triangular lattices. For the logarithmic scaling, the error bars are small and are not clearly visible.}
\label{Figure: ScalingAnalysis}
\end{figure*}

\begin{table}[b]
\caption{\label{Table: Constants and Exponents}
Estimated constants $A$ and $B$ in Eq.~\eqref{Equation: LogarithmicScaling}, exponents $a$ and $b$ in Eq.~\eqref{Equation: PowerLawScaling}, and reciprocal of the correlation length exponent $\nu^{'}$ associated with the specific heat for the hexagonal, square and triangular lattices. All of these constants and exponents are estimated from the same simulation data.}
\begin{ruledtabular}
\begin{tabular}{cccccc}
Lattice & \(A\) & \multicolumn{1}{c}{\(B\)} & \multicolumn{1}{c}{\(a\)} & \multicolumn{1}{c}{\(b\)} & \multicolumn{1}{c}{$\nu^{'}$} \\
\hline
Hexagonal & 0.4879(69) & 0.1695(49) & 0.94(12) & 1.03(3) & 1.014(24)\\
Square & 0.654(6) & 0.217(3) & 0.927(2) & 0.908(5) & 1.05(9)\\
Triangular & 0.928(13) & 0.2428(19) & 1.01(19) & 1.00(9) & 0.93(4)\\
\end{tabular}
\end{ruledtabular}
\end{table}
\end{widetext}

\bibliography{Bibliography}




\end{document}